\documentclass[preprint,12pt]{elsarticle}

\usepackage{amssymb}

\journal{XXX}

\begin{document}

\begin{frontmatter}



\title{Construction and enumeration for self-dual cyclic codes of even length
over $\mathbb{F}_{2^m} + u\mathbb{F}_{2^m}$}


\author{Yuan Cao$^{a, \ b, \ c}$, Yonglin Cao$^{a, \ \ast}$, Hai Q. Dinh$^{d, \ e}$, Fang-Wei Fu$^{f}$,
Fanghui Ma$^{f}$}

\address{$^{a}$School of Mathematics and Statistics,
Shandong University of Technology, Zibo, Shandong 255091, China\\

\vskip 1mm $^{b}$Hubei Key Laboratory of Applied Mathematics, Faculty of Mathematics and Statistics, Hubei University, Wuhan 430062, China

\vskip 1mm $^{c}$Hunan Provincial Key Laboratory of Mathematical Modeling and Analysis in Engineering, Changsha University of Science and Technology, Changsha, Hunan 410114, China

\vskip 1mm $^{d}$Division of Computational Mathematics and Engineering, Institute for Computational
       Science, Ton Duc Thang University, Ho Chi Minh City, Vietnam

\vskip 1mm $^{e}$Faculty of Mathematics and Statistics, Ton Duc Thang University, Ho Chi Minh City,
      Vietnam

\vskip 1mm $^{f}$Chern Institute of Mathematics and LPMC, Nankai University, Tianjin 300071, China
  }

\cortext[cor1]{corresponding author.  \\
E-mail addresses: yuancao@sdut.edu.cn (Yuan Cao), \ ylcao@sdut.edu.cn (Yonglin Cao),
\ dinhquanghai@tdtu.edu.vn (H. Q. Dinh), \ \ fwfu@nankai.edu.cn (F-W. Fu), \ fhma@mail.nankai.edu.cn (F. Ma).}

\begin{abstract}
Let $\mathbb{F}_{2^m}$ be a finite field of cardinality $2^m$, $R=\mathbb{F}_{2^m}+u\mathbb{F}_{2^m}$
$(u^2=0)$ and $s,n$ be positive integers such that $n$ is odd. In
this paper, we give an explicit representation for every self-dual cyclic code over the finite chain ring $R$
of length $2^sn$ and provide a calculation method to obtain all distinct codes.
Moreover, we obtain a clear formula to count the number of all these self-dual cyclic codes.
As an application, self-dual and $2$-quasi-cyclic codes over $\mathbb{F}_{2^m}$ of length $2^{s+1}n$ can be obtained
from self-dual cyclic code over $R$
of length $2^sn$ and by a Gray map preserving orthogonality and distances from $R$ onto $\mathbb{F}_{2^m}^2$.
\end{abstract}


\end{frontmatter}


\section{Introduction}

\noindent
  The class of self-dual codes is an interesting topic in coding theory duo to
their connections to other fields of mathematics such as Lattices, Cryptography, Invariant Theory, Block designs, etc.
A common theme for the construction of self-dual codes is the use of a computer search. In order to make this search feasible, special construction methods have been used to reduce the search field.
In many instances, self-dual codes have been found by first finding a code over a ring and then mapping
this code onto a code over a subring (subfield) through a map that preserves duality. In the literatures, the mappings typically map
to codes over $\mathbb{F}_2$, $\mathbb{F}_4$ and $\mathbb{Z}_4$ since codes over these rings have had the most use
(cf. \cite{s3}, \cite{s4}, \cite{s19}--\cite{s27}).

\par
   Let $\mathbb{F}_{p^m}$ be a finite field of $p^m$ elements, where $p$ is a prime number, and denote
$R=\frac{\mathbb{F}_{p^m}[u]}{\langle u^2\rangle}=\mathbb{F}_{p^m}
+u\mathbb{F}_{p^m} \ (u^2=0).$
Then $R$ is a finite chain ring and every invertible element in $R$ is of the form: $a+bu$, where $a,b\in \mathbb{F}_{p^m}$ and $a\neq 0$.
Let $N$ be a fixed positive integer, and denote
$R^N=\{(a_0,a_1,\ldots,a_{N-1})\mid a_0,a_1,\ldots,a_{N-1}\in R\}$
which is an $R$-free module with the usual componentwise addition and scalar multiplication by elements of $R$.
In coding theory, an
$R$-submodule of $R^N$ is called a \textit{linear code} over
$R$ of length $N$.

\par
  The \textit{Euclidean inner
product} on $R^N$ is defined by $[\alpha,\beta]=\sum_{i=0}^{N-1}a_ib_i\in R$ for
all vectors $\alpha=(a_0,a_1,\ldots,a_{N-1}), \beta=(b_0,b_1,\ldots,b_{N-1})\in R^N$. Then
the (Euclidean) \textit{dual code} of a linear code $\mathcal{C}$ over $R$ of length $N$ is defined by
$\mathcal{C}^{\bot}=\{\beta\in R^N\mid [\alpha,\beta]=0, \ \forall \alpha\in \mathcal{C}\}$,
which is also a linear code over $R$ of length $N$. In particular, $\mathcal{C}$ is said to be
 (Euclidean) \textit{self-dual} if $\mathcal{C}^{\bot}=\mathcal{C}$.

\par
  Let $\frac{R[x]}{\langle x^N-1\rangle}=\{\sum_{i=0}^{N-1}a_ix^i\mid a_0,a_1,\ldots,a_{N-1}\in R\}$
in which the arithmetic is done modulo $x^N-1$.
As usual, in this paper we identify \textit{cyclic codes} over
$R$ of length $N$ with ideals of the ring $\frac{R[x]}{\langle x^N-1\rangle}$ under the
identification map $\theta: R^N \rightarrow \frac{R[x]}{\langle x^N-1\rangle}$ defined by
$\theta: (a_0,a_1,\ldots,a_{N-1})\mapsto
a_0+a_1x+\ldots+a_{N-1}x^{N-1}$ for all $a_i\in R$.
Moreover,
$\mathcal{C}$ is called a \textit{simple-root cyclic code} if
${\rm gcd}(q,N)=1$, and called a \textit{repeated-root cyclic code} otherwise.

\par
  Let $p=2$ and $\alpha=a+bu\in R$ where $a,b\in \mathbb{F}_{2^m}$. As in \cite{s3}, we define
$\phi(\alpha)=(b,a+b)$ and define the \textit{Lee weight} of $\alpha$ by
${\rm w}_L(\alpha)={\rm w}_H(b,a+b)$, where ${\rm w}_H(b,a+b)$ is the Hamming weight of the vector
$(b,a+b)\in \mathbb{F}_{2^m}^2$. Then $\phi$ is an isomorphism of $\mathbb{F}_{2^m}$-linear
spaces from $R$ onto $\mathbb{F}_{2^m}^2$, and can be extended to an isomorphism of $\mathbb{F}_{2^m}$-linear
spaces from $\frac{R[x]}{\langle x^{N}-1\rangle}$ onto $\mathbb{F}_{2^m}^{2N}$ by:
$$
\phi(\xi)=(b_0,b_1,\ldots,b_{N-1},a_0+b_0,a_1+b_1,\ldots,a_{N-1}+b_{N-1}),
$$
for all $\xi=\sum_{i=0}^{N-1}\alpha_ix^i\in \frac{R[x]}{\langle x^{N}-1\rangle}$, where $\alpha_i=a_i+b_iu$ with $a_i,b_i\in \mathbb{F}_{2^m}$ and $i=0,1,\ldots,N-1$.

\par
  The following conclusion is derived from Corollary 14 of \cite{s3}:
   \textit{Let $\mathcal{C}$ be an ideal of $\frac{R[x]}{\langle x^{N}-1\rangle}$ and set $\phi(\mathcal{C})=\{\phi(\xi)\mid \xi\in \mathcal{C}\}\subseteq \mathbb{F}_{2^m}^{2N}$. Then}
\begin{description}
\item{(i)}
   \textit{$\phi(\mathcal{C})$ is a $2$-quasi-cyclic code over $\mathbb{F}_{2^m}$ of length $2N$}.

\item{(ii)}
 \textit{The Hamming weight distribution of $\phi(\mathcal{C})$ is exactly the same as the Lee weight distribution of $\mathcal{C}$}.

\item{(iii)}
 \textit{$\phi(\mathcal{C})$ is a self-dual code over $\mathbb{F}_{2^m}$ of length $2N$ if
$\mathcal{C}$ is a self-dual code over $R$ of length $N$}.
\end{description}
\noindent
 Hence it is an effective way to obtain self-dual and $2$-quasi-cyclic codes over $\mathbb{F}_{2^m}$ of length $2N$
from self-dual cyclic codes over $R$ of length $N$.

\par
  There were a lot of literatures on linear codes, cyclic codes and
constacyclic codes of length $N$ over rings $\mathbb{F}_{p^m}+u\mathbb{F}_{p^m}$ ($u^2=0$) for various prime $p$, positive integers $m$ and some positive integer $N$.
For example, \cite{s4}, \cite{s5}, \cite{s8}, \cite{s11}, \cite{s13}--\cite{s27}.
The classification of self-dual codes plays an important role in studying their structures and encoders.
However, it is a very difficult task in general, and only some codes of special lengths over certain finite
fields or finite chain rings are classified.

\par
  For example,
all constacyclic codes of length $2^s$ over the Galois extension
rings of $\mathbb{F}_2 + u\mathbb{F}_2$ was classified and their detailed structures was established in \cite{s13}. Dinh \cite{s14}
classified all constacyclic codes of length $p^s$ over $\mathbb{F}_{p^m}+u\mathbb{F}_{p^m}$.
    Dinh et al. \cite{s15} studied
negacyclic codes of length $2p^s$ over $\mathbb{F}_{p^m}+u\mathbb{F}_{p^m}$.
Chen et al. \cite{s11} investigated
constacyclic codes of length $2p^s$ over $\mathbb{F}_{p^m}+u\mathbb{F}_{p^m}$.
Dinh et al. \cite{s16} studied constacyclic codes of length $4p^s$ over $\mathbb{F}_{p^m}+u\mathbb{F}_{p^m}$ when $p^m\equiv 1$ (mod $4$).
These papers mainly used the methods in \cite{s13} and \cite{s14}, and the main results and their proofs  depend heavily on the code lengths $p^s$, $2p^s$ and $4p^s$. It is particularly important to note that \textsf{the representation and enumeration for self-dual cyclic codes were not studied in these papers}.

\par
   Dinh et al. \cite{s17} determined the algebraic structures of all cyclic and negacyclic codes
of length $4p^s$ over $\mathbb{F}_{p^m}+u\mathbb{F}_{p^m}$, established the duals of all such codes and gave some special subclass of self-dual negacyclic codes of length $4p^s$ over $\mathbb{F}_{p^m}+u\mathbb{F}_{p^m}$. But \textsf{the representation and enumeration for all self-dual negacyclic codes and all self-dual cyclic codes
were not obtained}.

\par
   Chen et al. \cite{s12} gave some new necessary and sufficient conditions for the existence of nontrivial self-dual \textsl{simple-root cyclic codes} over finite commutative chain rings and studied explicit enumeration formulas for these codes, but
\textsf{self-dual repeated-root cyclic codes over finite commutative chain rings were not considered}.

\par
  In \cite{s8} we gave an explicit representation for every self-dual cyclic code over $\mathbb{F}_{2^m}+u\mathbb{F}_{2^m}$ of length $2^s$
and obtained an exact formula to count the number of all these self-dual cyclic codes. Especially, we
provided an efficient method for the construction of all distinct self-dual cyclic codes with length $2^s$ over  $\mathbb{F}_{2^m}+u\mathbb{F}_{2^m}$ by use of properties for Kronecker product of matrices and calculation for linear equations over $\mathbb{F}_{2^m}$.

\par
   Recently, in \cite{s5} we provided a new way different
from the methods used in \cite{s11} and \cite{s13}--\cite{s18} to study
$\lambda$-constacyclic codes over $\mathbb{F}_{p^m}+u\mathbb{F}_{p^m}$ of length
$np^s$, where $\lambda\in \mathbb{F}_{p^m}^\times$ arbitrary and $n$ is an arbitrary positive integer satisfying ${\rm gcd}(p,n)=1$.
In particular, we obtained the following:

$\diamond$
Determined the algebraic structure and
 generators for each code. On that basis, we obtained many clear enumeration results for all codes.

$\diamond$
 Gave an explicit representation for the dual code of each $\lambda$-constacyclic
code over $\mathbb{F}_{p^m}+u\mathbb{F}_{p^m}$ of length
$np^s$.

$\diamond$
 Provided a clear distinguish condition (criteria) for the (Euclidean) self-duality
of each cyclic code and negacyclic code (corresponds to $\lambda=1$ and $\lambda=-1$, respectively).

\par
  Based on this, we consider further to give an explicit representation and enumeration for self-dual cyclic
codes and self-dual negacyclic codes over $\mathbb{F}_{p^m}+u\mathbb{F}_{p^m}$. In this paper, we focus on the case $p=2$.

\vskip 2mm\par
  The present paper is organized as follows.
In Section 2, we provide the necessary notation and preparatory conclusions.
Based on this, we give an explicit representation for all distinct self-dual cyclic codes of length $2^sn$ over $\mathbb{F}_{2^m}+u\mathbb{F}_{2^m}$ by Theorem 2.6. At the end, we obtain a clear
formula to count the number of these self-dual cyclic codes. In most cases, the
representation for each code in Theorem 2.6 is strongly dependent to determine
a special kind of subsets $\Omega_{j,\nu}$ in the
residue class ring $\frac{\mathbb{F}_{2^m}[x]}{\langle f_j(x)^\nu\rangle}$, for all irreducible
and self-reciprocal divisor $f_j(x)$ of $x^n-1$ in $\mathbb{F}_{2^m}[x]$ with degree $d_j>1$ and integers $\nu$,
$1\leq \nu\leq 2^{s-1}$. In Section 3, we give an effective algorithm to calculate these sets $\Omega_{j,\nu}$
and obtain a precise representation for them by use of trace functions.
In Section 4, we provide a proof for Theorem 2.6. In Section 5,
we give a way to calculate the number of self-dual cyclic codes over $\mathbb{F}_{2^m}+u\mathbb{F}_{2^m}$ of length
$2^sn$ from the positive integers $n, s, m$ directly. As an application, we list all $589$ self-dual cyclic codes over $\mathbb{F}_{2}+u\mathbb{F}_{2}$
of length $24$ precisely.
Section 6 concludes the paper.



\section{Explicit representation for self-dual cyclic codes over $R$}

\noindent
In this section, we give the result for representing and enumerating
all distinct self-dual cyclic codes of length $2^sn$
over $\mathbb{F}_{2^m} + u\mathbb{F}_{2^m}$ ($u^2=0$).
To do this, we introduce the necessary notation and conclusions first.

\par
  As $n$ is odd, there are distinct monic irreducible polynomials
$f_1(x)=x-1, f_2(x),\ldots,f_r(x)$ in $\mathbb{F}_{2^m}[x]$ such that
$x^n-1=\prod_{j=1}^rf_j(x)$. This implies
$$
x^{2^sn}-1=(x^n-1)^{2^s}=f_1(x)^{2^s}f_2(x)^{2^s}\ldots f_r(x)^{2^s}.$$
For any integer $j$, $1\leq j\leq r$, we assume ${\rm deg}(f_j(x))=d_j$ and denote $F_j(x)=\frac{x^{n}-1}{f_j(x)}$.
Then $F_j(x)^{2^s}=\frac{x^{2^sn}-1}{f_j(x)^{2^s}}$ and ${\rm gcd}(F_j(x),f_j(x))=1$. Hence there exist $v_j(x),w_j(x)\in \mathbb{F}_{2^m}[x]$ such that
 $v_j(x)F_j(x)+w_j(x)f_j(x)=1$. This implies
$$
v_j(x)^{2^s}F_j(x)^{2^s}+w_j(x)^{2^s}f_j(x)^{2^s}
=(v_j(x)F_j(x)+w_j(x)f_j(x))^{2^s}=1.
$$

\noindent
  In this paper, we adopt the following notations, where $1\leq j\leq r$.

\begin{itemize}
\item
  $\mathcal{A}=\frac{\mathbb{F}_{2^m}[x]}{\langle x^{2^sn}-1\rangle}=\{\sum_{i=0}^{2^sn-1}a_ix^i\mid
a_i\in \mathbb{F}_{2^m}, \ i=0,1,\ldots,2^sn-1\}$ in which the arithmetic is done modulo $x^{2^sn}-1$.

\item
   Let $\varepsilon_j(x)\in \mathcal{A}$ be defined by
$\varepsilon_j(x)\equiv v_j(x)^{2^s}F_j(x)^{2^s}=1-w_j(x)^{2^s}f_j(x)^{2^s}$ \ (mod $x^{2^sn}-1$).

Then $\sum_{j=1}^r\varepsilon_j(x)=1$, $\varepsilon_j(x)^2=\varepsilon_j(x)$ and $\varepsilon_j(x)\varepsilon_l(x)=0$ for all $j\neq l$.

\item
  $\mathcal{K}_j=\frac{\mathbb{F}_{2^m}[x]}{\langle f_j(x)^{2^s}\rangle}=\{\sum_{i=0}^{2^sd_j-1}a_ix^i\mid
a_i\in \mathbb{F}_{2^m}, \ i=0,1,\ldots,2^sd_j-1\}$ in which the arithmetic is done modulo $f_j(x)^{2^s}$.

\item
  $\mathcal{F}_j=\frac{\mathbb{F}_{2^m}[x]}{\langle f_j(x)\rangle}=\{\sum_{i=0}^{d_j-1}a_ix^i\mid
a_i\in \mathbb{F}_{2^m}, \ i=0,1,\ldots, d_j-1\}$ in which the arithmetic is done modulo $f_j(x)$. It is
well known that $\mathcal{F}_j$ is an extension field of $\mathbb{F}_{2^m}$ with degree $d_j$, and hence
$|\mathcal{F}_j|=2^{d_jm}$.

\item
   $\frac{\mathcal{A}[u]}{\langle u^2\rangle}=\mathcal{A}+u\mathcal{A}=\{a(x)+ub(x)\mid a(x),b(x)\in \mathcal{A}\}$ ($u^2=0$).

\item
   $\frac{\mathcal{K}_j[u]}{\langle u^2\rangle}=\mathcal{K}_j+u\mathcal{K}_j=\{a(x)+ub(x)\mid a(x),b(x)\in \mathcal{K}_j\}$ ($u^2=0$).   \end{itemize}

\noindent
  {\bf Remark} $\mathcal{F}_j$ is a finite field with operations defined by the usual polynomial operations
modulo $f_j(x)$, $\mathcal{K}_j$ is a finite chain ring with operations defined by the usual polynomial operations
modulo $f_j(x)^{2^s}$ and $\mathcal{A}$ is a principal ideal ring with operations defined by the usual polynomial operations
modulo $x^{2^sn}-1$. In this paper, we adopt the following points of view:
$$\mathcal{F}_j\subseteq \mathcal{K}_j\subseteq \mathcal{A} \ {\rm and}
\ \mathcal{K}_j+u\mathcal{K}_j\subseteq \mathcal{A}+u\mathcal{A} \ {\rm as} \ {\rm sets}.$$
Obviously, $\mathcal{F}_j$ is not a subfield of $\mathcal{K}_j$, $\mathcal{K}_j$ is not a subring of $\mathcal{A}$
and $\mathcal{K}_j+u \mathcal{K}_j$ is not a subring of $\mathcal{A}+u\mathcal{A}$ when
$n\geq 2$.

\vskip 3mm\par
  Now, we consider how to determine cyclic codes over $\mathbb{F}_{2^m}+u\mathbb{F}_{2^m}$ of
length $2^sn$, i.e. ideals of the ring $\frac{(\mathbb{F}_{2^m}+u\mathbb{F}_{2^m})[x]}{\langle x^{2^sn}-1\rangle}$.

\vskip 3mm\noindent
  {\bf Lemma 2.1} (cf. \cite{s5} Lemma 2.2)
 \textit{For any $\xi=a(x)+ub(x)\in \mathcal{A}+u\mathcal{A}$, where $a(x)=\sum_{j=0}^{2^sn-1}a_jx^j,
b(x)=\sum_{j=0}^{2^sn-1}b_jx^j\in \mathcal{A}$ with $a_j,b_j\in \mathbb{F}_{2^m}$, we define
$$\Psi(\xi)=a(x)+ub(x)=\sum_{0\leq j\leq 2^sn-1}(a_j+ub_j)x^j.$$
Then $\Psi$ is a ring isomorphism from $\mathcal{A}+u\mathcal{A}$ onto $\frac{(\mathbb{F}_{2^m}+u\mathbb{F}_{2^m})[x]}{\langle x^{2^sn}-1\rangle}$}.

\vskip 3mm\par
   From now on, we will identify $\mathcal{A}+u\mathcal{A}$ with $\frac{(\mathbb{F}_{2^m}+u\mathbb{F}_{2^m})[x]}{\langle x^{2^sn}-1\rangle}$
under the ring isomorphism $\Psi$ defined in Lemma 2.1. Then
in order to determine all distinct cyclic codes over $\mathbb{F}_{2^m}+u\mathbb{F}_{2^m}$ of length $2^sn$,
it is sufficient to give all distinct
ideals of the ring $\mathcal{A}+u\mathcal{A}$. For the latter, we have the following conclusion.

\vskip 3mm\noindent
  {\bf Lemma 2.2}
   (cf. \cite{s5} Theorem 2.7 and Corollary 3.9)
 \textit{Let $\mathcal{C}\subseteq\mathcal{A}+u\mathcal{A}$. Then $\mathcal{C}$ is an ideal
of $\mathcal{A}+u\mathcal{A}$ if and only if for each integer $j$, $1\leq j\leq r$,
there is a unique ideal $C_j$ of the ring $\mathcal{K}_j+u\mathcal{K}_j$ such that
$$\mathcal{C}=\varepsilon_1(x)C_1\oplus\varepsilon_2(x)C_2
\oplus\ldots\oplus \varepsilon_r(x)C_r \ ({\rm mod} \ x^{2^sn}-1),$$
where $\varepsilon_j(x)C_j=\{\varepsilon_j(x)c_j(x)\mid c_j(x)\in C_j\}$ for all
$j=1,\ldots,r$. In this case, the number of codewords in $\mathcal{C}$ is
$|\mathcal{C}|=\prod_{j=1}^r|C_j|$}.

\vskip 3mm\par
  All distinct ideals of $\mathcal{K}_j+u\mathcal{K}_j$ had been listed
by Theorem 3.8 of [5] for all $j$. Here, we give
the structure of its subring $\mathcal{K}_j$.

\vskip 3mm\noindent
  {\bf Lemma 2.3}
  (cf. \cite{s6} Lemma 3.7 and \cite{s7} Example 2.1) \textit{The ring $\mathcal{K}_j$ have the following properties}:

\begin{description}
\item{(i)}
  \textit{$\mathcal{K}_j$ is a finite chain ring, $f_j(x)$ generates the unique
maximal ideal $\langle f_j(x)\rangle$ of $\mathcal{K}_j$, $2^s$ is the nilpotency index of $f_j(x)$ and the residue class field of $\mathcal{K}_j$ modulo $\langle f_j(x)\rangle$ is $\mathcal{K}_j/\langle f_j(x)\rangle\cong \mathcal{F}_{j}$}.

\item{(ii)}
 \textit{We regard $\mathcal{F}_j$ as a subset of $\mathcal{K}_j$ in the sense of Remark before Lemma 2.1.
Then  every element $\xi$ of $\mathcal{K}_j$ has a unique $f_j(x)$-adic expansion}: \\
$\xi=b_0(x)+b_1(x)f_j(x)+\ldots+ b_{2^s-1}f_j(x)^{2^s-1}, \ b_0(x), b_1(x),\ldots, b_{2^s-1}\in \mathcal{F}_j.$

\item{(iii)}
   \textit{All distinct ideals of $\mathcal{K}_j$ are given by: $\langle f_j(x)^l\rangle=f_j(x)^l\mathcal{K}_j$, $0\leq l\leq 2^s$. Moreover, $|\langle f_j(x)^l\rangle|=2^{md_j(2^s-l)}$ for $l=0,1,\ldots,2^s$}.

\item{(iv)}
  \textit{Let $1\leq l\leq 2^s$. Then
$\mathcal{K}_j/\langle f_j(x)^l\rangle=\{\sum_{k=0}^{l-1}b_k(x)f(x)^k\mid b_k(x)\in \mathcal{F}_j, \ k=0,1,\ldots,l-1\}$,}
\textit{and hence $|\mathcal{K}_j/\langle f_j(x)^l\rangle|=2^{md_jl}$}.

\item{(v)}
  \textit{For any $0\leq l\leq t\leq 2^s-1$, we have
$$f_j(x)^l(\mathcal{K}_j/\langle f_j(x)^t\rangle)=\{\sum_{k=l}^{t-1}b_k(x)f(x)^k\mid
b_k(x)\in \mathcal{F}_j, \ k=l,\ldots,t-1\}$$
and $|f_j(x)^l(\mathcal{K}_j/\langle f_j(x)^t\rangle)|=2^{md_j(t-l)}$, where we set
$f_j(x)^l(\mathcal{K}_j/\langle f_j(x)^l\rangle)$ $=\{0\}$ for convenience}.
\end{description}

\noindent
  {\bf Remark} For any integer $l$, $1\leq l\leq 2^s-1$, by Lemma 2.3(iv)
we can identify $\mathcal{K}_j/\langle f_j(x)^l\rangle$ with $\frac{\mathbb{F}_{2^m}[x]}{\langle f_j(x)^l\rangle}$ up to a natural
ring isomorphism. We will take this view in the rest of this paper.
  Then for any
$0\leq l\leq t\leq 2^s-1$, we stipulate
$$f_j(x)^l\cdot \frac{\mathbb{F}_{2^m}[x]}{\langle f_j(x)^t\rangle}=f_j(x)^l(\mathcal{K}_j/\langle f_j(x)^t\rangle).$$

\par
  For any polynomial $f(x)=\sum_{i=0}^dc_ix^i\in \mathbb{F}_{2^m}[x]$ of degree $d\geq 1$,
the \textit{reciprocal polynomial} of $f(x)$ is defined as $\widetilde{f}(x)=\widetilde{f(x)}=x^df(\frac{1}{x})=\sum_{i=0}^dc_ix^{d-i}$.
 $f(x)$ is said to be \textit{self-reciprocal} if $\widetilde{f}(x)=\delta f(x)$ for some $\delta\in \mathbb{F}_{2^m}^{\times}$.
It is known that $\widetilde{\widetilde{f}(x)}=f(x)$ if $f(0)\neq 0$, and $\widetilde{f(x)g(x)}=\widetilde{f}(x)\widetilde{g}(x)$ for
any polynomials $f(x), g(x)\in\mathbb{F}_{2^m}[x]$ with positive degrees satisfying $f(0),g(0)\in \mathbb{F}_{2^m}^{\times}$.

\par
   As
$x^n-1=\prod_{j=1}^rf_j(x)$, where $f_1(x)=x-1, f_2(x),\ldots, f_r(x)$
are pairwise coprime monic irreducible polynomials in $\mathbb{F}_{2^m}[x]$, it follows that
$$x^n-1=x^n+1=\widetilde{(x^n+1)}
=\widetilde{f}_1(x)\widetilde{f}_2(x)\ldots\widetilde{f}_r(x),$$
where $\widetilde{f}_1(x)=f_1(x), \widetilde{f}_2(x), \ldots,\widetilde{f}_r(x)$
are pairwise coprime monic irreducible polynomials in $\mathbb{F}_{2^m}[x]$ as well. Hence after a rearrangement of $f_2(x),\ldots, f_r(x)$, there are integers $\rho\geq 1$ and $\epsilon\geq 0$ such that
\begin{description}
\item{$\diamond$}
   $r=\rho+2\epsilon$;

\item{$\diamond$}
   $\widetilde{f}_j(x)=\delta_jf_j(x)$ for some $\delta_j\in \mathbb{F}_{2^m}^{\times}$, for all $j=1,\ldots,\rho$;

\item{$\diamond$}
 $\widetilde{f}_j(x)=\delta_jf_{j+\epsilon}(x)$ for some $\delta_j\in \mathbb{F}_{2^m}^{\times}$, for all $j=\rho+1,\ldots,\rho+\epsilon$.
\end{description}

\noindent
 The following lemma can be easily verified from the definition of self-reciprocal
 polynomials over $\mathbb{F}_{2^m}$.

\vskip 3mm\noindent
 {\bf Lemma 2.4} (cf. \cite{s9} Lemma 3.2) \textit{Using the notation above,
we have the following conclusions, where $d_j={\rm deg}(f_j(x))$}.

\begin{description}
\item{(i)}
  \textit{$\delta_j=1$ and $\widetilde{f}_j(x)=f_j(x)$, for all $j=1,\ldots,\rho$}.

\item{(ii)}
  \textit{$d_j$ is even for all $j=2,\ldots,\rho$}.
\end{description}

\par
   Let $A=(a_{ij})$ and $B$ be matrices over $\mathbb{F}_{2^m}$ of sizes $s\times t$ and $l\times v$ respectively.
Recall that the \textit{Kronecker product} of $A$ and $B$ is
defined by $A\otimes B=(a_{ij}B)$ which is a matrix over $\mathbb{F}_{2^m}$ of size $sl\times tv$. Then we denote
$$G_2=\left(\begin{array}{cc} 1 & 0 \cr 1 & 1\end{array}\right),
\ G_{2^\lambda}=G_2\otimes G_{2^{\lambda-1}}=\left(\begin{array}{cc} G_{2^{\lambda-1}} & 0 \cr G_{2^{\lambda-1}} & G_{2^{\lambda-1}}\end{array}\right),
\ \lambda=2,3,\ldots.$$
Denote by $I_{2^\lambda}$ the
identity matrix of order $2^\lambda$. For any $2\leq l\leq 2^s-1$, let $M_l$ be the
submatrix in the upper left corner of
$I_{2^\lambda}+G_{2^\lambda}$, i.e.,
\begin{itemize}
\item
$\left(\begin{array}{cc} M_l & 0 \cr \ast & \ast\end{array}\right)=I_{2^\lambda}+G_{2^\lambda}$,
where $M_l$ is a matrix over $\mathbb{F}_{2}$ of size $l\times l$.
\end{itemize}
Especially, we have $M_{2^\lambda}=I_{2^\lambda}+G_{2^\lambda}$.

\par
  For any matrix $A$ over $\mathbb{F}_{2^m}$, let $A^{{\rm tr}}$ be the transpose of $A$. In the rest of this paper, we adopt
the following notation.
\begin{itemize}
\item
$\mathcal{S}_l=\{B_l=(b_0,b_1,\ldots,b_{l-1})^{{\rm tr}} \mid M_lB_l=0, \
b_0,b_1,\ldots,b_{l-1}\in \mathbb{F}_{2^m}\}.$

\item
$\mathcal{S}_l^{[\delta]}=\{B_l^{[\delta]}=(b_\delta\, \ldots, b_{l-1})^{{\rm tr}}\mid
(0,\ldots,0,b_\delta\, \ldots, b_{l-1})^{{\rm tr}} \in\mathcal{S}_l\}$,  $0\leq \delta<l$. \\
Then $\mathcal{S}_1^{[0]}=\mathbb{F}_{2^m}$ and $\mathcal{S}_3^{[1]}=\mathbb{F}_{2^m}^2$ (cf. \cite{s8} Theorem 2(i)).

\item
  For any integers $j$ and $\nu$, $2\leq j\leq \rho$ and $1\leq \nu\leq 2^{s-1}$, denote
$$\Omega_{j,\nu}=\left\{\beta(x)\in \frac{\mathbb{F}_{2^m}[x]}{\langle f_j(x)^\nu\rangle}\mid \beta(x)+x^{-\nu d_j}\beta(x^{-1})\equiv 0 \ ({\rm mod} \ f_j(x)^\nu)\right\}.$$
\end{itemize}

\par
  We will give an effective algorithm to determine the subset $\Omega_{j,\nu}$ of $\frac{\mathbb{F}_{2^m}[x]}{\langle f_j(x)^\nu\rangle}$ and prove that $|\Omega_{j,\nu}|=2^{\nu m\frac{d_j}{2}}$ in the next section of this paper.

  A recursive algorithm to determine the subspace $\mathcal{S}_l$ of $\mathbb{F}_{2^m}^l$ was given
by Theorem 1 in \cite{s8}. Moreover, we have the following conclusion for $\mathcal{S}_l^{[\delta]}$.

\vskip 3mm\noindent
 {\bf Lemma 2.5} (cf. \cite{s8} Theorem 2(i))
 \textit{Using the notations above, let $s\geq 3$ and $1\leq h\leq 2^{s-2}-1$. Then}.

\begin{description}
\item{$\diamond$}
 $\mathcal{S}_{2^s-1}^{[2^{s-1}-1]}=\{(b_{2^{s-1}-1},b_{2^{s-1}},\ldots, b_{2^s-3}, b_{2^s-2})^{{\rm tr}}\mid b_{2^{s-2}+2i-1}, b_{2^s-2}\in \mathbb{F}_{2^m}$, $i=0,1,\ldots,2^{s-2}-1\}$
   \textit{in which
 $b_{2^{s-1}+2i}$ is a fixed $\mathbb{F}_{2^m}$-linear combination of $b_{2^{s-1}+2z-1}$, $0\leq z\leq i$, for all $i=0,1,\ldots,2^{s-2}-2$}.

\item{$\diamond$}
  $\mathcal{S}_{4h-1}^{[2h-1]}
=\{(b_{2h-1},b_{2h},b_{2h+1},\ldots,b_{4h-3}$, $b_{4h-2})^{{\rm tr}}\mid b_{2h+2i-1}\in \mathbb{F}_{2^m}$,
 $i=0,1,\ldots,h-1, \ {\rm and} \ b_{4h-2}\in \mathbb{F}_{2^m}\}$
 \textit{in which
 $b_{2h+2i}$ is a fixed $\mathbb{F}_{2^m}$-linear combination of $b_{2h+2z-1}$, $0\leq z\leq i$, for all $i=0,1,\ldots,h-2$}.

\item{$\diamond$}
 $\mathcal{S}_{4h+1}^{[2h]}
=\{(0,b_{2h+1},\ldots,b_{4h-1}$, $b_{4h})^{{\rm tr}}\mid b_{2h+1+2i}\in \mathbb{F}_{2^m}, \
 i=0,1,\ldots,h-1, \ {\rm and} \ b_{4h}\in \mathbb{F}_{2^m}\}$
 \textit{in which
 $b_{2h+2+2i}$ is a fixed $\mathbb{F}_{2^m}$-linear combination of $b_{2h+2+2z-1}$, $0\leq z\leq i$, for all $i=0,1,\ldots, h-2$}.

\item{$\diamond$}
 $|\mathcal{S}_{2^s-1}^{[2^{s-1}-1]}|=(2^m)^{2^{s-2}+1}$ and $|\mathcal{S}_{4h-1}^{[2h-1]}|=|\mathcal{S}_{4h+1}^{[2h]}|=(2^m)^{h+1}$.
\end{description}

\par
  Now is the time to list self-dual cyclic codes of length $2^sn$
over $\mathbb{F}_{2^m} + u\mathbb{F}_{2^m}$.

\vskip 3mm
\noindent
  {\bf Theorem 2.6}
\textit{Using the notations above, all distinct
self-dual cyclic codes of length $2^sn$ over $\mathbb{F}_{2^m}+u\mathbb{F}_{2^m}$ are given by
$$\mathcal{C}=\bigoplus_{j=1}^r\varepsilon_j(x)C_j \ ({\rm mod} \ x^{2^sn}-1),$$
where $C_j$ is an ideal of $\mathcal{K}_j+u\mathcal{K}_j$ given by the following three cases}:

\begin{description}
\item{(I)}
   \textit{$C_1$ is given by one of the following three subcases}.

\begin{description}
\item{$\triangleright$} \textit{If $s=1$, there are $N_1=1+2^m$ ideals}:
\begin{description}
\item{}
   $C_1=\langle x+1\rangle$;
   \textit{$C_1=\langle (x+1)b+u\rangle$ where $b\in \mathbb{F}_{2^m}$}.
\end{description}

\item{$\triangleright$} \textit{If $s=2$, there are $N_2=1+2^m+(2^m)^2$
ideals}:

\begin{description}
\item{}
   $C_1=\langle (x+1)^2\rangle$;

\vskip 2mm\item{}
   \textit{$C_1=\langle (x+1)b(x)+u\rangle$, \\
   where $b(x)=b_1(x+1)+b_2(x+1)^2$ with $b_1,b_2\in \mathbb{F}_{2^m}$};

\vskip 2mm\item{}
   \textit{$C_1=\langle (x+1)^2b+u(x+1),(x+1)^3\rangle$ where $b\in \mathbb{F}_{2^m}$}.
\end{description}

\item{$\triangleright$} \textit{If $s\geq 3$, there are
$N_s=1+2^m+2(2^m)^2\cdot\frac{(2^m)^{2^{s-2}-1}-1}{2^m-1}+(2^m)^{2^{s-2}+1}$
ideals}:

\begin{description}
\item{}
  $C_1=\langle (x+1)^{2^{s-1}}\rangle$.

\vskip 2mm\item{}
  \textit{$C_1=\langle (x+1)b(x)+u\rangle$, where $b(x)=\sum_{i=2^{s-1}-1}^{2^s-2}b_i(x+1)^i$
with $(b_{2^{s-1}-1},b_{2^{s-1}}$, $\ldots,b_{2^s-2})^{{\rm tr}}\in \mathcal{S}_{2^s-1}^{[2^{s-1}-1]}$}.

\vskip 2mm\item{}
  \textit{$C_1=\langle (x+1)^{2^{s-1}}b+u(x+1)^{2^{s-1}-1},(x+1)^{2^{s-1}+1}\rangle$ where $b\in \mathbb{F}_{2^m}$}.

\vskip 2mm\item{}
  \textit{$C_1=\langle (x+1)^{2^{s-1}-2h+1}b(x)+u(x+1)^{2^{s-1}-2h},(x+1)^{2^{s-1}+2h}\rangle$, \\
  where $b(x)=\sum_{i=2h-1}^{4h-2}b_i(x+1)^i$, $1\leq h\leq 2^{s-2}-1$ and \\
  $(b_{2h-1},b_{2h},b_{2h+1},\ldots, b_{4h-3},b_{4h-2})^{{\rm tr}}\in \mathcal{S}_{4h-1}^{[2h-1]}$}.

\vskip 2mm\item{}
  \textit{$C_1=\langle (x+1)^{2^{s-1}-2h}b(x)+u(x+1)^{2^{s-1}-2h-1},(x+1)^{2^{s-1}+2h+1}\rangle$, where
  $b(x)=\sum_{i=2h+1}^{4h}b_i(x+1)^i$, $1\leq h\leq 2^{s-2}-1$ and \\
  $(0,b_{2h+1},b_{2h+2},\ldots, b_{4h-1},b_{4h})^{{\rm tr}}\in \mathcal{S}_{4h+1}^{[2h]}$}.
\end{description}
\end{description}

\item{(II)}
   \textit{If $2\leq j\leq \rho$, $C_j$ is one of the following $1+\sum_{\nu=1}^{2^{s-1}}\left(2^{m\frac{d_j}{2}}\right)^\nu$
ideals}:

\begin{description}
\item{(ii-1)}
   $C_j=\langle f_j(x)^{2^{s-1}}\rangle$.

\vskip 2mm
\item{(ii-2)}
   \textit{$C_j=\langle f_j(x)b(x)+u\rangle$, where
$b(x)\in f_j(x)^{2^{s-1}-1}\cdot \Omega_{j,2^{s-1}}$}.

\vskip 2mm
\item{(ii-3)}
    \textit{$C_j=\langle f_j(x)^{2^{s-1}-\nu+1}b(x)+uf_j(x)^{2^{s-1}-\nu},f_j(x)^{2^{s-1}+\nu}\rangle$, \\
     where $b(x)\in f_j(x)^{\nu-1}\cdot \Omega_{j,\nu}$ and $1\leq \nu\leq 2^{s-1}-1$}.
\end{description}

\item{(III)}
   \textit{If $j=\rho+i$ where $1\leq i\leq \epsilon$, there are
   $$N_{(2^m,d_j,2^{s})}=\sum_{i=0}^{2^{s-1}}(1+4i)2^{(2^{s-1}-i)md_j}$$
   pairs $(C_j,C_{j+\epsilon})$ of ideals given by the following five subcases,
  where $x^{-1}=x^{2^sn-1}$ (mod $f_{j+\epsilon}(x)^{2^s}$)}:

\begin{description}
\item{(iii-1)}
   \textit{$C_j=\langle f_j(x)b(x)+u\rangle$ and $C_{j+\epsilon}=\langle \delta_jx^{2^sn-d_j}f_{j+\epsilon}(x)b(x^{-1})+u\rangle$
  where
$b(x)\in f_j(x)^{2^{s-1}-1}\cdot \frac{\mathbb{F}_{2^m}[x]}{\langle f_j(x)^{2^s-1}\rangle}$}.

\item{(iii-2)}
   \textit{$C_j=\langle f_j(x)^{k+1}b(x)+uf_j(x)^k\rangle$ and
\begin{center}
$C_{j+\epsilon}=\langle \delta_j x^{2^sn-d_j}f_{j+\epsilon}(x)b(x^{-1})+u, f_{j+\epsilon}(x)^{2^s-k}\rangle$,
\end{center}
where $b(x)\in f_j(x)^{\lceil \frac{2^s-k}{2}\rceil-1}\cdot \frac{\mathbb{F}_{2^m}[x]}{\langle f_j(x)^{2^s-k-1}\rangle}$
 and $1\leq k\leq 2^s-1$}.

\item{(iii-3)}
  \textit{$C_j=\langle f_j(x)^k\rangle$ and $C_{j+\epsilon}=\langle f_{j+\epsilon}(x)^{2^s-k}\rangle$, where $0\leq k\leq 2^s$}.

\item{(iii-4)}
   \textit{$C_j=\langle f_j(x)b(x)+u,f_j(x)^{t}\rangle$ and
\begin{center}
$C_{j+\epsilon}=\langle \delta_j x^{2^sn-d_j}f_{j+\epsilon}(x)^{2^s-t+1}b(x^{-1})+uf_{j+\epsilon}(x)^{2^s-t}\rangle$,
\end{center}
where $b(x)\in f_j(x)^{\lceil\frac{t}{2}\rceil-1}\cdot \frac{\mathbb{F}_{2^m}[x]}{\langle f_j(x)^{t-1}\rangle}$ and $1\leq t\leq 2^s-1$}.

\item{(iii-5)}
   \textit{$C_j=\langle f_j(x)^{k+1}b(x)+uf_j(x)^k,f_j(x)^{k+t}\rangle$ and
\begin{eqnarray*}
C_{j+\epsilon}&=&\langle \delta_j x^{2^sn-d_j}f_{j+\epsilon}(x)^{2^s-k-t+1}b(x^{-1})+uf_{j+\epsilon}(x)^{2^s-k-t},\\
&&f_{j+\epsilon}(x)^{2^s-k}\rangle,
\end{eqnarray*}
where $b(x)\in f_j(x)^{\lceil\frac{t}{2}\rceil-1}\cdot \frac{\mathbb{F}_{2^m}[x]}{\langle f_j(x)^{t-1}\rangle}$, $1\leq t\leq 2^s-k-1$ and \\
$1\leq k\leq 2^s-2$}.
\end{description}
\end{description}

\par
  Finally, from Theorem 2.6 we obtain a mass formula to count the number of self-dual cyclic codes over $\mathbb{F}_{2^m}+u\mathbb{F}_{2^m}$ of length $2^sn$
for any integer $s\geq 2$.

\vskip 3mm\noindent
 {\bf Theorem 2.7} \textit{Using the notation above, we have the following conclusions}.

\begin{description}
\item{(i)}
  \textit{Let $s\geq 3$. Then the number ${\rm NE}(2^m,2^sn)$ of self-dual cyclic codes over $\mathbb{F}_{2^m}+u\mathbb{F}_{2^m}$ of length $2^sn$
  is equal to}
\begin{eqnarray*}
 &&\left(1+2^m+2(2^m)^2\cdot\frac{(2^m)^{2^{s-2}-1}-1}{2^m-1}+(2^m)^{2^{s-2}+1}\right)\\
 &&\cdot \prod_{j=2}^\rho\left(1+\sum_{\nu=1}^{2^{s-1}}\left(2^{m\frac{d_j}{2}}\right)^\nu\right)  \cdot\prod_{j=\rho+1}^{\rho+\epsilon}\left(\sum_{i=0}^{2^{s-1}}(1+4i)2^{(2^{s-1}-i)md_j}\right).
\end{eqnarray*}

\item{(ii)} \textit{When $s=2$, the number ${\rm NE}(2^m,4n)$ of self-dual cyclic codes over $\mathbb{F}_{2^m}+u\mathbb{F}_{2^m}$ of length $4n$
  is equal to}
  $$(1+2^m+4^m)\cdot \prod_{j=2}^\rho(1+2^{m\frac{d_j}{2}}+2^{md_j})
  \cdot\prod_{j=\rho+1}^{\rho+\epsilon}(9+5\cdot 2^{md_j}+4^{md_j}).$$
\end{description}

\par
  For any odd positive integer $n$, the number of self-dual cyclic codes over $\mathbb{F}_{2^m}+u\mathbb{F}_{2^m}$ of length $2n$
had been determined by Corollary 4.1 in \cite{s9}.



\section{Determining the subset $\Omega_{j,\nu}$ of $\frac{\mathbb{F}_{2^m}[x]}{\langle f_j(x)^\nu\rangle}$}
\noindent
In this section, we solve the key problem for determining self-dual cyclic codes over $\mathbb{F}_{2^m}+u\mathbb{F}_{2^m}$ of length $2^sn$ in Theorem 2.6:
  \textsf{Provide an effective algorithm to determine the subset $\Omega_{j,\nu}$ of $\frac{\mathbb{F}_{2^m}[x]}{\langle f_j(x)^\nu\rangle}$},
for all $2\leq j\leq \rho$ and
$1\leq \nu\leq 2^{s-1}$.  To do this, we introduce the following notation:

\begin{itemize}
\item
   $\mathcal{H}_j=\{\xi\in \mathcal{F}_j\mid \xi^{2^{m\frac{d_j}{2}}}=\xi\}.$
Is is well known that $\mathcal{H}_j$ is a subfield of $\mathcal{F}_j=\frac{\mathbb{F}_{2^m}[x]}{\langle f_j(x)\rangle}$
with $2^{m\frac{d_j}{2}}$ elements (cf. \cite{s29} Chapter 6).

\item
   Let
${\rm Tr}_j$ be the trace function from $\mathcal{F}_j$ onto $\mathcal{H}_j$ defined by: \\
${\rm Tr}_j(\xi)=\xi+\xi^{2^{m\frac{d_j}{2}}}\in \mathcal{H}_j \ (\forall \xi\in \mathcal{F}_j).$

\item
   For any $\alpha\in \mathcal{H}_j$, denote by
${\rm Tr}_j^{-1}(\alpha)=\{\xi\in \mathcal{F}_j\mid {\rm Tr}_j(\xi)=\alpha\}$
the pre-image of $\alpha$ in the $\mathcal{F}_j$. Then it is well
known that
$|{\rm Tr}_j^{-1}(\alpha)|=2^{\frac{d_j}{2}m}$ (cf. \cite{s29} Corollary 7.17).
\end{itemize}

\noindent
  {\bf Lemma 3.1} \textit{Let $2\leq j\leq \rho$ and $1\leq \nu\leq 2^{s-1}$. Using the notation above and in Section 2, we have the following conclusions}.

\begin{description}
\item{(i)}
  \textit{Let $\zeta_j(x)$ be a primitive element of the finite field
$\mathcal{F}_j=\frac{\mathbb{F}_{2^m}[x]}{\langle f_j(x)\rangle}$. Then} \\
${\rm Tr}_j^{-1}(0)=\mathcal{H}_j=\{0\}\cup\{\zeta_j(x)^{(2^{m\frac{d_j}{2}}+1)l}\mid l=0,1,\ldots,2^{m\frac{d_j}{2}}-2\}.$
\item{(ii)}
  \textit{Let $\alpha\in \mathcal{H}_j$ and $\widehat{\alpha}\in \mathcal{F}_j$
satisfying ${\rm Tr}_j(\widehat{\alpha})=\widehat{\alpha}+\widehat{\alpha}^{2^{m\frac{d_j}{2}}}=\alpha$. Then} \\
${\rm Tr}_j^{-1}(\alpha)=\{\widehat{\alpha}\}\cup\{\widehat{\alpha}+\zeta_j(x)^{(2^{m\frac{d_j}{2}}+1)l}\mid l=0,1,\ldots,2^{m\frac{d_j}{2}}-2\}.$
\end{description}

\noindent
  {\bf Proof.} (i) Let $\eta_j(x)=\zeta_{j}(x)^{2^{m\frac{d_j}{2}}+1}\in \mathcal{F}_j$. Then
by ${\rm ord}(\zeta_{j}(x))=2^{md_j}-1=(2^{m\frac{d_j}{2}}+1)(2^{m\frac{d_j}{2}}-1)$, it follows
that ${\rm ord}(\eta_{j}(x))=2^{m\frac{d_j}{2}}-1$. Since $\mathcal{H}_j$ is the unique
subfield of $\mathcal{F}_j$ with $2^{m\frac{d_j}{2}}$ elements, we see that
$\eta_{j}(x)$ is a primitive element of $\mathcal{H}_j$. This implies that
$\mathcal{H}_j=\{0\}\cup\{\eta_j(x)^l\mid l=0,1,\ldots,2^{m\frac{d_j}{2}}-2\}$.

\par
  For any $\alpha\in \mathcal{H}_j$, we have $\alpha^{2^{m\frac{d_j}{2}}}=\alpha$. This implies
${\rm Tr}_j(\alpha)=\alpha+\alpha^{2^{m\frac{d_j}{2}}}=0$. Hence $\mathcal{H}_j\subseteq {\rm Tr}_j^{-1}(0)$. From this and
by $|{\rm Tr}_j^{-1}(0)|=\frac{|\mathcal{F}_j|}{|\mathcal{H}_j|}=2^{m\frac{d_j}{2}}=|\mathcal{H}_j|$, we deduce
that ${\rm Tr}_j^{-1}(0)=\mathcal{H}_j$.

\par
  (ii) For any $\gamma\in \mathcal{H}_j$, by ${\rm Tr}_j(\widehat{\alpha})=\alpha$ it follows that
${\rm Tr}_j(\widehat{\alpha}+\gamma)={\rm Tr}_j(\widehat{\alpha})+{\rm Tr}_j(\gamma)=\alpha+0=\alpha$, and
hence $\widehat{\alpha}+\mathcal{H}_j\subseteq {\rm Tr}_j^{-1}(\alpha)$. From this,  by
$|\widehat{\alpha}+\mathcal{H}_j|=|\mathcal{H}_j|=2^{m\frac{d_j}{2}}=|{\rm Tr}_j^{-1}(\alpha)|$ and (i) we deduce
that ${\rm Tr}_j^{-1}(\alpha)=\widehat{\alpha}+\mathcal{H}_j=\{\widehat{\alpha}\}\cup\{\widehat{\alpha}+\zeta_j(x)^{(2^{m\frac{d_j}{2}}+1)l}\mid l=0,1,\ldots,2^{m\frac{d_j}{2}}-2\}$.
\hfill
$\Box$

\vskip 3mm\par
  Now, the set $\Omega_{j,\nu}$ can be determined by the following theorem.

\vskip 3mm \noindent
  {\bf Theorem 3.2} \textit{Assume $2\leq j\leq \rho$ and $1\leq \nu\leq 2^{s-1}$}.
  \textit{Let
  $$\beta(x)=\sum_{0\leq i\leq \nu-1}\beta_i(x)f_j(x)^i\in \Omega_{j,\nu} \
  {\rm arbitrary},$$
where $\beta_0(x),\beta_1(x),\ldots,\beta_{\nu-1}(x)\in \mathcal{F}_j$. Then $\beta_0(x),\beta_1(x),\ldots,\beta_{\nu-1}(x)$ can be determined by the following processes, where $x^{-1}=x^{2^sn-1}$ $({\rm mod} \ f_j(x)^\nu)$}:
\begin{description}
\item{{\bf Step 0}.}
  \textit{Let $\mathcal{W}^{(j,\nu; 0)}=x^{-\nu \frac{d_j}{2}}\cdot {\rm Tr}_j^{-1}(0)$.}
  \textit{Then $|\mathcal{W}^{(j,\nu; 0)}|=2^{m\frac{d_j}{2}}$ and $\beta_0(x)\in \mathcal{W}^{(j,\nu; 0)}$}.

\item{{\bf Step 1}.} \textit{Let $\beta_0(x)\in \mathcal{W}^{(j,\nu; 0)}$. Then there is a unique
sequence}:
$$\delta^{(1)}_{(\beta_0)}(x),\delta^{(2)}_{(\beta_0)}(x),\ldots,\delta^{(\nu-1)}_{(\beta_0)}(x)\in \mathcal{F}_j$$
\textit{satisfying}
$\beta_{0}(x)+x^{-(\nu+0)d_j}\beta_{0}(x^{-1})=\sum_{i=1}^{\nu-1}\delta^{(i)}_{(\beta_0)}(x)f_j(x)^i
\ ({\rm mod} \ f_j(x)^\nu).$
\textit{Moreover, we have that $x^{(\nu+i)\frac{d_j}{2}}\delta^{(i)}_{(\beta_0)}(x)\in \mathcal{H}_j$ for all $i=1,2,\ldots,\nu-1$}.

$\bullet$ \textit{Set}
$\mathcal{W}^{(j,\nu; 1)}_{(\beta_0)}=x^{-(\nu+1)\frac{d_j}{2}}\cdot
{\rm Tr}_j^{-1}\left(x^{(\nu+1)\frac{d_j}{2}}\delta^{(1)}_{(\beta_0)}(x)\right).$

  \textit{Then $|\mathcal{W}^{(j,\nu; 1)}_{(\beta_0)}|=2^{m\frac{d_j}{2}}$ and $\beta_1(x)\in \mathcal{W}^{(j,\nu; 1)}_{(\beta_0)}$}.

\item{{\bf Step 2}.} \textit{Let $\beta_0(x)\in \mathcal{W}^{(j,\nu; 0)}$
and $\beta_1(x)\in \mathcal{W}^{(j,\nu; 1)}_{(\beta_0)}$. Then there is a unique
sequence
$$\delta^{(2)}_{(\beta_0,\beta_1)}(x),\delta^{(3)}_{(\beta_0,\beta_1)}(x),\ldots,\delta^{(\nu-1)}_{(\beta_0,\beta_1)}(x)\in \mathcal{F}_j$$
satisfying}
$$\beta_{1}(x)+x^{-(\nu+1)d_j}\beta_{1}(x^{-1})+\delta_{(\beta_0)}^{(1)}
  =\sum_{i=2}^{\nu-1}\delta^{(i)}_{(\beta_0,\beta_1)}(x)f_j(x)^{i-1}
  \ ({\rm mod} \ f_j(x)^{\nu-1}).$$
\textit{Moreover, we have that $x^{(\nu+i)\frac{d_j}{2}}\delta^{(i)}_{(\beta_0,\beta_1)}(x)\in \mathcal{H}_j$ for all $i=2,3,\ldots,\nu-1$}.

  $\bullet$ \textit{Set}
$\mathcal{W}^{(j,\nu; 2)}_{(\beta_0,\beta_1)}=x^{-(\nu+2)\frac{d_j}{2}}\cdot
{\rm Tr}_j^{-1}\left(x^{(\nu+2)\frac{d_j}{2}}\left(\delta^{(2)}_{(\beta_0)}(x)+\delta^{(2)}_{(\beta_0,\beta_1)}(x)\right)\right).$

  \textit{Then $|\mathcal{W}^{(j,\nu; 2)}_{(\beta_0,\beta_1)}|=2^{m\frac{d_j}{2}}$ and $\beta_2(x)\in \mathcal{W}^{(j,\nu; 2)}_{(\beta_0,\beta_1)}$}.

\item{{\bf Step i}.} \textit{For any integer $i=2,\ldots,\nu-1$, where $\nu\geq 3$,
let $\beta_0(x)\in \mathcal{W}^{(j,\nu; 0)}, \beta_1(x)\in \mathcal{W}^{(j,\nu; 1)}_{(\beta_0)},
\beta_2(x)\in \mathcal{W}^{(j,\nu; 2)}_{(\beta_0,\beta_1)}, \ldots,\beta_{i-1}(x)\in \mathcal{W}^{(i-1)}_{(\beta_0,\beta_1,\ldots,\beta_{i-2})}$. Then there is a unique
sequence}
$$\delta^{(i)}_{(\beta_0,\beta_1,\ldots,\beta_{i-2},\beta_{i-1})}(x),
\delta^{(i+1)}_{(\beta_0,\beta_1,\ldots,\beta_{i-2},\beta_{i-1})}(x),\ldots,
\delta^{(\nu-1)}_{(\beta_0,\beta_1,\ldots,\beta_{i-2},\beta_{i-1})}(x)\in \mathcal{F}_j$$
\textit{satisfying}
\begin{eqnarray*}
&&\beta_{i-1}(x)+x^{-(\nu+i-1)d_j}\beta_{i-1}(x^{-1})+\sum_{h=0}^{i-2}\delta^{(i-1)}_{(\beta_0,\beta_1,\ldots,\beta_{h})}(x)\\
&=&\sum_{l=i}^{\nu-1}\delta^{(l)}_{(\beta_0,\beta_1,\ldots,\beta_{i-2},\beta_{i-1})}(x)f_j(x)^{l-i+1}
\ ({\rm mod} \ f_j(x)^{\nu-i+1}).
\end{eqnarray*}
\textit{Moreover, we have that $x^{(\nu+l)\frac{d_j}{2}}\delta^{(l)}_{(\beta_0,\beta_1,\ldots,\beta_{i-2},\beta_{i-1})}(x)\in \mathcal{H}_j$ for all $i=i,i+1,\ldots,\nu-1$}.

$\bullet$ \textit{Set}
$$\mathcal{W}^{(j,\nu; i)}_{(\beta_0,\beta_1,\ldots,\beta_{i-1})}
=x^{-(\nu+i)\frac{d_j}{2}}\cdot
{\rm Tr}_j^{-1}\left(x^{(\nu+i)\frac{d_j}{2}}
\left(\sum_{h=0}^{i-1}\delta^{(i)}_{(\beta_0,\beta_1,\ldots,\beta_{h})}(x)\right)\right).$$
  \textit{Then $|\mathcal{W}^{(j,\nu; i)}_{(\beta_0,\beta_1,\ldots,\beta_{i-1})}|=2^{m\frac{d_j}{2}}$ and $\beta_i(x)\in \mathcal{W}^{(j,\nu; i)}_{(\beta_0,\beta_1,\ldots,\beta_{i-1})}$}.
\end{description}

  \textit{Therefore, we have}
$$\Omega_{j,\nu}=\left\{\sum_{i=0}^{\nu-1}\beta_i(x)f_j(x)^i\mid \beta_i(x)\in \mathcal{W}^{(j,\nu; i)}_{(\beta_0,\ldots,\beta_{i-2},\beta_{i-1})}, \ 0\leq i\leq\nu-1\right\}$$
\textit{in which we set $\mathcal{W}^{(j,\nu; i)}_{(\beta_0,\ldots,\beta_{i-2},\beta_{i-1})}=\mathcal{W}^{(j,\nu; 0)}$ if
$i=0$}.

\textit{Hence the number of elements in $\Omega_{j,\nu}$ is $|\Omega_{j,\nu}|=2^{\nu m\frac{d_j}{2}}$}.

\vskip 3mm \noindent
  {\bf Proof.} As $2\leq j\leq \lambda$, $d_j$ is an even positive integer by Lemma 2.4(ii). It is well known that
$x^{-1}=x^{2^{m\frac{d_j}{2}}}$ in the finite field $\mathcal{F}_j=\frac{\mathbb{F}_{2^m}[x]}{\langle f_j(x)\rangle}$.
From this and by $a^{2^m}=a$ in $\mathbb{F}_{2^m}$, we deduce that
\begin{equation}
\label{eq1}
\alpha(x)^{2^{m\frac{d_j}{2}}}=\alpha(x^{2^{m\frac{d_j}{2}}})=\alpha(x^{-1}),
\ \forall \alpha(x)\in \mathcal{F}_j.
\end{equation}

\par
  Let $\beta(x)=\sum_{i=0}^{\nu-1}\beta_i(x)f_j(x)^i\in
  \Omega_{j,\nu}\subset \frac{\mathbb{F}_{2^m}[x]}{\langle f_j(x)^\nu\rangle}$, where
$\beta_i(x) \in \mathcal{F}_j$ for all $i=0,1,\ldots,\nu-1$. As $\widetilde{f}_j(x)=f_j(x)$ by Lemma 2.4(i), we have
$$f_j(x^{-1})=x^{-d_j}(x^{d_j}f_j(x^{-1}))=x^{-d_j}\widetilde{f}_j(x)=x^{-d_j}f_j(x) \ {\rm in} \ \frac{\mathbb{F}_{2^m}[x]}{\langle f_j(x)^\nu\rangle}.$$
This implies
$$\beta(x^{-1})=\sum_{i=0}^{\nu-1}\beta_i(x^{-1})f_j(x^{-1})^i=\sum_{i=0}^{\nu-1}x^{-id_j}\beta_i(x^{-1})f_j(x)^i
 \ {\rm in} \ \frac{\mathbb{F}_{2^m}[x]}{\langle f_j(x)^\nu\rangle},$$
and
$\beta(x)+x^{-\nu d_j}\beta(x^{-1})\equiv \sum_{i=0}^{\nu-1}(\beta_i(x)+x^{-(\nu+i)d_j}\beta_i(x^{-1}))f_j(x)^i$
(mod $f_j(x)^\nu$). From this and by the definition of $\Omega_{j,\nu}$, we deduce that
\begin{equation}
\label{eq2}
\sum_{i=0}^{\nu-1}\left(\beta_i(x)+x^{-(\nu+i)d_j}\beta_i(x^{-1})\right)f_j(x)^i\equiv 0
\ ({\rm mod} \ f_j(x)^\nu).
\end{equation}

\vskip 2mm\par
   {\bf Step 0.} We prove that $\beta_0(x)\in \mathcal{W}^{(j,\nu; 0)}$. In fact,
by Equation (\ref{eq2}) we have
$$\beta_0(x)+x^{-(\nu+0)d_j}\beta_0(x^{-1})\equiv 0 \  ({\rm mod} \ f_j(x)),$$
i.e., $\beta_0(x)+x^{-\nu d_j}\beta_0(x^{-1})=0$ in
$\mathcal{F}_j$.
Both sides multiplied by $x^{\nu\frac{d_j}{2}}$, the above equation is equivalent to
$$x^{\nu\frac{d_j}{2}}\beta_0(x)+(x^{\nu\frac{d_j}{2}}\beta_0(x))^{2^{m\frac{d_j}{2}}}
=x^{\nu\frac{d_j}{2}}\beta_0(x)+x^{-\nu\frac{d_j}{2}}\beta_0(x^{-1})=0  \ {\rm in} \ \mathcal{F}_j,$$
i.e., ${\rm Tr}_j(x^{\nu\frac{d_j}{2}}\beta_0(x))=0$. Hence $x^{\nu\frac{d_j}{2}}\beta_0(x)\in {\rm Tr}_j^{-1}(0)$,
this is equivalent to
$$\beta_0(x)=x^{-\nu\frac{d_j}{2}}(x^{\nu \frac{d_j}{2}}\beta_0(x))\in x^{-\nu\frac{d_j}{2}}\cdot {\rm Tr}_j^{-1}(0)
=\mathcal{W}^{(j,\nu; 0)}.$$
Obviously, $|\mathcal{W}^{(j,\nu; 0)}|=|{\rm Tr}_j^{-1}(0)|=2^{m\frac{d_j}{2}}$.

\vskip 2mm\par
  {\bf Step 1.} Let $\beta_0(x)\in \mathcal{W}^{(j,\nu; 0)}$ where $\nu\geq 2$.
Then $\beta_0(x)+x^{-\nu d_j}\beta_0(x^{-1})\equiv 0$ (mod $f_j(x)$).
Hence there exist
polynomials $\delta_{(\beta_0)}^{(1)}(x), \delta_{(\beta_0)}^{(2)}(x),\ldots,\delta_{(\beta_0)}^{(\nu-1)}(x)\in \mathcal{F}_j$
satisfying
\begin{equation}
\label{eq3}
\beta_{0}(x)+x^{-\nu d_j}\beta_{0}(x^{-1})=\sum_{i=1}^{\nu-1}\delta^{(i)}_{(\beta_0)}(x)f_j(x)^i
\ ({\rm mod} \ f_j(x)^\nu).
\end{equation}
From this, we deduce $\delta_{(\beta_0)}^{(1)}(x)=\frac{\beta_0(x)+x^{-\nu d_j}\beta_0(x^{-1})}{f_j(x)}$
(mod $f_j(x)$). By $f_j(x^{-1})=x^{-d_j}f_j(x)$, it follows that
$$x^{\frac{d_j}{2}}f_j(x^{-1})=x^{-\frac{d_j}{2}}f_j(x).$$
This implies $x^{-\frac{d_j}{2}}f_j(x)\in \mathcal{H}_j$. From this and by $x^{(\nu+1)\frac{d_j}{2}}=\frac{x^{\nu \frac{d_j}{2}}}{x^{-\frac{d_j}{2}}}$,
we obtain
$$x^{(\nu+1)\frac{d_j}{2}}\delta_{(\beta_0)}^{(1)}(x)=\frac{x^{\nu \frac{d_j}{2}}\beta_0(x)+x^{-\nu \frac{d_j}{2}}\beta_0(x^{-1})}{x^{-\frac{d_j}{2}}f_j(x)}
\ ({\rm mod} \ f_j(x)).$$
This implies that $\left(x^{(\nu+1)\frac{d_j}{2}}\delta_{(\beta_0)}^{(1)}(x)\right)^{2^{m\frac{d_j}{2}}}
=x^{(\nu+1)\frac{d_j}{2}}\delta_{(\beta_0)}^{(1)}(x)$ (mod $f_j(x)$) by Equation (\ref{eq1}),
and hence $x^{(\nu+1)\frac{d_j}{2}}\delta_{(\beta_0)}^{(1)}(x)\in \mathcal{H}_j$.

\par
  Now, let $2\leq i\leq \nu-1$, where $\nu\geq 3$,
and assume that $x^{(\nu+1)\frac{d_j}{2}}\delta_{(\beta_0)}^{(1)}(x),\ldots$,
$x^{(\nu+i-1)\frac{d_j}{2}}\delta_{(\beta_0)}^{(i-1)}(x)$ $\in \mathcal{H}_j$. By Equation (\ref{eq3}), we have
$$\delta_{(\beta_0)}^{(i)}(x)=\frac{\beta_0(x)+x^{-\nu d_j}\beta_0(x^{-1})
+\sum_{l=1}^{i-1}\delta_{(\beta_0)}^{(l)}(x)f_j(x)^l}{f_j(x)^i}
\ ({\rm mod} \ f_j(x)).$$
From this, by
$x^{(\nu+i)\frac{d_j}{2}}=\frac{x^{\nu \frac{d_j}{2}}}{\left(x^{-\frac{d_j}{2}}\right)^i}
\ {\rm and} \ x^{\nu \frac{d_j}{2}}=x^{(\nu+l) \frac{d_j}{2}}\cdot \left(x^{-\frac{d_j}{2}}\right)^l$
we obtain
\begin{eqnarray*}
&&x^{(\nu+i)\frac{d_j}{2}}\delta_{(\beta_0)}^{(i)}(x)\\
&=&\frac{x^{\nu \frac{d_j}{2}}\beta_0(x)+x^{-\nu \frac{d_j}{2}}\beta_0(x^{-1})
+\sum_{l=1}^{i-1}\left(x^{(\nu+l) \frac{d_j}{2}}\delta_{(\beta_0)}^{(l)}(x)\right) \left(x^{-\frac{d_j}{2}}f_j(x)\right)^l}
{\left(x^{-\frac{d_j}{2}}f_j(x)\right)^i}.
\end{eqnarray*}
This implies that $\left(x^{(\nu+i)\frac{d_j}{2}}\delta_{(\beta_0)}^{(i)}(x)\right)^{2^{m\frac{d_j}{2}}}
=x^{(\nu+i)\frac{d_j}{2}}\delta_{(\beta_0)}^{(i)}(x)$ (mod $f_j(x)$) by Equation (\ref{eq1}),
and hence $x^{(\nu+i)\frac{d_j}{2}}\delta_{(\beta_0)}^{(i)}(x)\in \mathcal{H}_j$.

\par
  According to the inductive principle, we conclude that
$x^{(\nu+i)\frac{d_j}{2}}\delta_{(\beta_0)}^{(i)}(x)\in \mathcal{H}_j$ for all $i=1,2,\ldots,\nu-1$.

\par
  Then by combining two Equations (\ref{eq2}) and (\ref{eq3}), we get
$$\sum_{i=1}^{\nu-1}\left(\beta_i(x)+x^{-(\nu+i)d_j}\beta_i(x^{-1})+\delta^{(i)}_{(\beta_0)}(x)\right)f_j(x)^{i}\equiv 0
\ ({\rm mod} \ f_j(x)^\nu).$$
This implies
\begin{equation}
\label{eq4}
\sum_{i=1}^{\nu-1}\left(\beta_i(x)+x^{-(\nu+i)d_j}\beta_i(x^{-1})+\delta^{(i)}_{(\beta_0)}(x)\right)f_j(x)^{i-1}\equiv 0
\ ({\rm mod} \ f_j(x)^{\nu-1}).
\end{equation}

\par
 Finally, by $x^{(\nu+1)\frac{d_j}{2}}\delta_{(\beta_0)}^{(1)}(x)\in \mathcal{H}_j$ we have
$|{\rm Tr}_j^{-1}\left(x^{(\nu+1)\frac{d_j}{2}}\delta_{(\beta_0)}^{(1)}(x)\right)|=2^{m\frac{d_j}{2}}$. Hence
$\mathcal{W}_{(\beta_0)}^{(j,\nu; 1)}=x^{-(\nu+1)\frac{d_j}{2}}\cdot {\rm Tr}_j^{-1}\left(x^{(\nu+1)\frac{d_j}{2}}\delta_{(\beta_0)}^{(1)}(x)\right)$
is well-defined and $|\mathcal{W}_{(\beta_0)}^{(j,\nu; 1)}|=2^{m\frac{d_j}{2}}$.
Furthermore, by Equation (\ref{eq4}) and $\nu\geq 2$ we have
$$\beta_1(x)+x^{-(\nu+1)d_j}\beta_1(x^{-1})+\delta^{(1)}_{(\beta_0)}(x)\equiv 0
\ ({\rm mod} \ f_j(x)),$$
which is equivalent to
$$x^{(\nu+1)\frac{d_j}{2}}\beta_1(x)+x^{-(\nu+1)\frac{d_j}{2}}\beta_1(x^{-1})=x^{(\nu+1)\frac{d_j}{2}}\delta^{(1)}_{(\beta_0)}(x)
\in \mathcal{H}_j.$$
This implies $x^{(\nu+1)\frac{d_j}{2}}\beta_1(x)\in {\rm Tr}_j^{-1}(x^{(\nu+1)\frac{d_j}{2}}\delta^{(1)}_{(\beta_0)}(x))$
by Equation (\ref{eq1}).
Therefore, $\beta_1(x)=x^{-(\nu+1)\frac{d_j}{2}}\left(x^{(\nu+1)\frac{d_j}{2}}\beta_1(x)\right)\in \mathcal{W}_{(\beta_0)}^{(j,\nu; 1)}$.

\vskip 2mm\par
  {\bf Step 2.} Let
$\beta_0(x)\in \mathcal{W}^{(j,\nu; 0)}$ and $\beta_1(x)\in \mathcal{W}^{(j,\nu; 1)}_{(\beta_0)}$
where $\nu\geq 3$. Then we have
$\beta_1(x)+x^{-(\nu+1)d_j}\beta_1(x^{-1})+\delta^{(1)}_{(\beta_0)}(x)\equiv 0$ (mod $f_j(x)$).
Hence there exist
polynomials $\delta_{(\beta_0,\beta_1)}^{(2)}(x), \delta_{(\beta_0,\beta_1)}^{(3)}(x),\ldots,\delta_{(\beta_0,\beta_1)}^{(\nu-1)}(x)\in \mathcal{F}_j$
satisfying
\begin{equation}
\label{eq5}
\beta_{1}(x)+x^{-(\nu+1)d_j}\beta_{1}(x^{-1})+\delta_{(\beta_0)}^{(1)}
  =\sum_{i=2}^{\nu-1}\delta^{(i)}_{(\beta_0,\beta_1)}(x)f_j(x)^{i-1}
  \ ({\rm mod} \ f_j(x)^{\nu-1}).
\end{equation}
From this, we deduce
$$\delta_{(\beta_0,\beta_1)}^{(i)}(x)=\frac{\beta_1(x)+x^{-(\nu+1) d_j}\beta_1(x^{-1})+\delta^{(1)}_{(\beta_0)}(x)
+\sum_{l=2}^{i-1}\delta_{(\beta_0,\beta_1)}^{(l)}(x)f_j(x)^{l-1}}{f_j(x)^{i-1}}$$
(mod $f_j(x))$) for all $i=2,3,\ldots,\nu-1$. Here we set $\sum_{l=2}^{i-1}\delta_{(\beta_0,\beta_1)}^{(l)}(x)f_j(x)^{l-1}=0$ if
$i=2$. Then by
$x^{(\nu+i)\frac{d_j}{2}}=\frac{x^{(\nu+1) \frac{d_j}{2}}}{\left(x^{-\frac{d_j}{2}}\right)^{i-1}}
\ {\rm and} \ x^{(\nu+1) \frac{d_j}{2}}=x^{(\nu+l) \frac{d_j}{2}}\cdot \left(x^{-\frac{d_j}{2}}\right)^{l-1}$,
we obtain
\begin{eqnarray*}
&&x^{(\nu+i)\frac{d_j}{2}}\delta_{(\beta_0,\beta_1)}^{(i)}(x)\\
&=&\frac{1}{\left(x^{-\frac{d_j}{2}}f_j(x)\right)^{i-1}}
   \left(x^{(\nu+1) \frac{d_j}{2}}\beta_1(x)+x^{-(\nu+1) \frac{d_j}{2}}\beta_1(x^{-1})
      +x^{(\nu+1) \frac{d_j}{2}}\delta^{(1)}_{(\beta_0)}(x)\right.\\
 &&\left.+\sum_{l=2}^{i-1}\left(x^{(\nu+l) \frac{d_j}{2}}\delta_{(\beta_0,\beta_1)}^{(l)}(x)\right)
  \left(x^{-\frac{d_j}{2}}f_j(x)\right)^{l-1}\right).
\end{eqnarray*}
From this, by Equation (\ref{eq1}) and according to the inductive assumptions
we deduce that
$x^{(\nu+i)\frac{d_j}{2}}\delta_{(\beta_0,\beta_1)}^{(i)}(x)\in \mathcal{H}_j$
for all $i=2,3,\ldots,\nu-1$.

\par
  Then by combining two formulas (\ref{eq4}) and (\ref{eq5}), we get
$$\sum_{i=2}^{\nu-1}\left(\beta_i(x)+x^{-(\nu+i)d_j}\beta_i(x^{-1})+\delta^{(i)}_{(\beta_0)}(x)
+\delta^{(i)}_{(\beta_0,\beta_1)}(x)\right)f_j(x)^{i}\equiv 0$$
(mod $f_j(x)^\nu$). This implies that
\begin{equation}
\label{eq6}
\sum_{i=2}^{\nu-1}\left(\beta_i(x)+x^{-(\nu+i)d_j}\beta_i(x^{-1})+\delta^{(i)}_{(\beta_0)}(x)
+\delta^{(i)}_{(\beta_0,\beta_1)}(x)\right)f_j(x)^{i-2}\equiv 0
\end{equation}
(mod $f_j(x)^{\nu-2}$). Since
$\mathcal{H}_j$ is a subfield of $\mathcal{F}_j$ and
$$x^{(\nu+2)\frac{d_j}{2}}\delta_{(\beta_0)}^{(2)}(x),
x^{(\nu+2)\frac{d_j}{2}}\delta_{(\beta_0,\beta_1)}^{(2)}(x)\in \mathcal{H}_j,$$ we have
$x^{(\nu+2)\frac{d_j}{2}}(\delta_{(\beta_0)}^{(2)}(x)+
\delta_{(\beta_0,\beta_1)}^{(2)}(x))\in \mathcal{H}_j$. Hence
$$\mathcal{W}_{(\beta_0,\beta_1)}^{(j,\nu; 2)}=x^{-(\nu+2)\frac{d_j}{2}}\cdot {\rm Tr}_j^{-1}\left(x^{(\nu+2)\frac{d_j}{2}}\left(\delta_{(\beta_0)}^{(2)}(x)+\delta_{(\beta_0,\beta_1)}^{(2)}(x)\right)\right)$$
is well-defined and $|\mathcal{W}_{(\beta_0)}^{(j,\nu; 1)}|=|{\rm Tr}_j^{-1}\left(x^{(\nu+2)\frac{d_j}{2}}\left(\delta_{(\beta_0)}^{(2)}(x)+\delta_{(\beta_0,\beta_1)}^{(2)}(x)\right)\right)|=2^{m\frac{d_j}{2}}$. Furthermore, by Equation (\ref{eq6}) we have
$$\beta_2(x)+x^{-(\nu+2)d_j}\beta_2(x^{-1})+\delta^{(2)}_{(\beta_0)}(x)+\delta^{(2)}_{(\beta_0,\beta_1)}(x)\equiv 0
\ ({\rm mod} \ f_j(x)),$$
which is equivalent to
$$x^{(\nu+2)\frac{d_j}{2}}\beta_2(x)+x^{-(\nu+2)\frac{d_j}{2}}\beta_2(x^{-1})=x^{(\nu+2)\frac{d_j}{2}}\left(\delta^{(2)}_{(\beta_0)}(x)
+\delta^{(2)}_{(\beta_0,\beta_1)}(x)\right)\in \mathcal{H}_j.$$
This implies $x^{(\nu+2)\frac{d_j}{2}}\beta_2(x)\in {\rm Tr}_j^{-1}\left(x^{(\nu+2)\frac{d_j}{2}}\left(\delta^{(2)}_{(\beta_0)}(x)
+\delta^{(2)}_{(\beta_0,\beta_1)}(x)\right)\right)$
by Equation (\ref{eq1}).
Therefore, $\beta_2(x)=x^{-(\nu+2)\frac{d_j}{2}}\left(x^{(\nu+2)\frac{d_j}{2}}\beta_2(x)\right)
\in \mathcal{W}_{(\beta_0,\beta_1)}^{(j,\nu; 2)}$.

\vskip 2mm
\par
  {\bf Step i.} Let $\nu\geq 4$ and $3\leq i\leq \nu-1$. Assume that
$\mathcal{W}^{(j,\nu; 0)}$, $\mathcal{W}_{(\beta_0)}^{(j,\nu; 1)}$, $\ldots$,
$\mathcal{W}^{(i-1)}_{(\beta_0\ldots,\beta_{i-2})}$ have been determined, and
let $\beta_{i-1}(x)\in \mathcal{W}^{(i-1)}_{(\beta_0,\beta_1,\ldots,\beta_{i-2})}$, where
$$\mathcal{W}^{(i-1)}_{(\beta_0,\beta_1,\ldots,\beta_{i-2})}=x^{-(\nu+i-1)\frac{d_j}{2}}\cdot
{\rm Tr}_j^{-1}\left(x^{(\nu+i-1)\frac{d_j}{2}}\sum_{h=0}^{i-2}\delta^{(i-1)}_{(\beta_0,\beta_1,\ldots,\beta_{h})}(x)\right).$$
Then we have
$$\beta_{i-1}(x)+x^{-(\nu+i-1)d_j}\beta_{i-1}(x^{-1})
+\sum_{h=0}^{i-2}\delta^{(i-1)}_{(\beta_0,\beta_1,\ldots,\beta_{h})}(x)\equiv 0
\ ({\rm mod} \ f_j(x)).$$
Hence there exist
polynomials $\delta_{(\beta_0,\beta_1,\ldots,\beta_{i-2},\beta_{i-1})}^{(l)}(x)\in \mathcal{F}_j$,
for all $l=i,i+1,\ldots,\nu-1$,
satisfying
$$\beta_{i-1}(x)+x^{-(\nu+i-1)d_j}\beta_{i-1}(x^{-1})+\sum_{h=0}^{i-2}\delta^{(i-1)}_{(\beta_0,\beta_1,\ldots,\beta_{h})}(x)$$
\begin{equation}
\label{eq7}
=\sum_{l=i}^{\nu-1}\delta^{(l)}_{(\beta_0,\beta_1,\ldots,\beta_{i-2},\beta_{i-1})}(x)f_j(x)^{l-i+1}
\ ({\rm mod} \ f_j(x)^{\nu-i+1}).
\end{equation}
From this, we deduce
\begin{eqnarray*}
&&\delta_{(\beta_0,\beta_1,\ldots,\beta_{i-2},\beta_{i-1})}^{(l)}(x) \\
&=&\frac{1}{f_j(x)^{l-i+1}}\left(\beta_{i-1}(x)+x^{-(\nu+i-1)d_j}\beta_{i-1}(x^{-1})
+\sum_{h=0}^{i-2}\delta^{(i-1)}_{(\beta_0,\beta_1,\ldots,\beta_{h})}(x)\right.\\
&&\left. +\sum_{z=i}^{l-1}\delta^{(z)}_{(\beta_0,\beta_1,\ldots,\beta_{i-2},\beta_{i-1})}(x)f_j(x)^{z-i+1}\right)
\end{eqnarray*}
for all $l=i,i+1,\ldots,\nu-1$. Here we set $\sum_{z=i}^{l-1}\delta^{(z)}_{(\beta_0,\beta_1,\ldots,\beta_{i-2},\beta_{i-1})}(x)f_j(x)^{z-i+1}=0$
if $l=i$. Then by
$$x^{(\nu+l)\frac{d_j}{2}}=\frac{x^{(\nu+i-1) \frac{d_j}{2}}}{\left(x^{-\frac{d_j}{2}}\right)^{l-i+1}}
\ {\rm and} \ x^{(\nu+i-1) \frac{d_j}{2}}=x^{(\nu+z) \frac{d_j}{2}}\cdot \left(x^{-\frac{d_j}{2}}\right)^{z-i+1}$$
we obtain
\begin{eqnarray*}
&&x^{(\nu+l)\frac{d_j}{2}}\delta_{(\beta_0,\beta_1,\ldots,\beta_{i-2},\beta_{i-1})}^{(l)}(x) \\
&=&\frac{1}{\left(x^{-\frac{d_j}{2}}f_j(x)\right)^{l-i+1}}
 \left(x^{(\nu+i-1) \frac{d_j}{2}}\beta_{i-1}(x)+x^{-(\nu+i-1)\frac{d_j}{2}}\beta_{i-1}(x^{-1})\right.\\
 &&+\sum_{h=0}^{i-2}x^{(\nu+i-1) \frac{d_j}{2}}\delta^{(i-1)}_{(\beta_0,\beta_1,\ldots,\beta_{h})}(x)\\
 &&\left. +\sum_{z=i}^{l-1}\left(x^{(\nu+z) \frac{d_j}{2}}\delta^{(z)}_{(\beta_0,\beta_1,\ldots,\beta_{i-2},\beta_{i-1})}(x)\right)
 \left(x^{-\frac{d_j}{2}}f_j(x)\right)^{z-i+1}\right).
\end{eqnarray*}
From this, based on inductive assumptions and by Equation (\ref{eq1})
we deduce that
$x^{(\nu+l)\frac{d_j}{2}}\delta_{(\beta_0,\beta_1,\ldots,\beta_{i-2},\beta_{i-1})}^{(l)}(x) \in \mathcal{H}_j$
for all $l=i,i+1,\ldots,\nu-1$.

\par
  Based on inductive assumptions and by Equation (\ref{eq7}), we get
$$\sum_{l=i}^{\nu-1}f_j(x)^l\left(\beta_l(x)+x^{-(\nu+l)d_j}\beta_l(x^{-1})
+\sum_{h=0}^{i-1}\delta^{(l)}_{(\beta_0,\beta_1,\ldots,\beta_{h})}(x)\right)\equiv 0$$
(mod $f_j(x)^\nu$). This implies that
$$\sum_{l=i}^{\nu-1}f_j(x)^{l-i}\left(\beta_l(x)+x^{-(\nu+l)d_j}\beta_l(x^{-1})
+\sum_{h=0}^{i-1}\delta^{(l)}_{(\beta_0,\beta_1,\ldots,\beta_{h})}(x)\right)\equiv 0$$
(mod $f_j(x)^{\nu-i}$), and hence
$$\beta_i(x)+x^{-(\nu+i)d_j}\beta_i(x^{-1})
+\sum_{h=0}^{i-1}\delta^{(i)}_{(\beta_0,\beta_1,\ldots,\beta_{h})}(x)\equiv 0
\ ({\rm mod} \ f_j(x)).$$
Then in the ring $\mathcal{F}_j$, we have
\begin{equation}
\label{eq8}
x^{(\nu+i)\frac{d_j}{2}}\beta_i(x)+x^{-(\nu+i)\frac{d_j}{2}}\beta_i(x^{-1})
\equiv x^{(\nu+i)\frac{d_j}{2}}\sum_{h=0}^{i-1}\delta^{(i)}_{(\beta_0,\beta_1,\ldots,\beta_{h})}(x)
\end{equation}
Based on inductive assumptions, we have $\sum_{h=0}^{i-1}x^{(\nu+i)\frac{d_j}{2}}\delta^{(i)}_{(\beta_0,\beta_1,\ldots,\beta_{h})}(x)\in \mathcal{H}_j$. Hence
$$\mathcal{W}_{(\beta_0,\beta_1,\ldots,\beta_{i-1})}^{(j,\nu; i)}=x^{-(\nu+i)\frac{d_j}{2}}\cdot {\rm Tr}_j^{-1}\left(x^{(\nu+i)\frac{d_j}{2}}
\sum_{h=0}^{i-1}\delta^{(i)}_{(\beta_0,\beta_1,\ldots,\beta_{h})}(x)\right)$$
is well-defined and $|\mathcal{W}_{(\beta_0,\beta_1,\ldots,\beta_{i-1})}^{(j,\nu; i)}|=2^{m\frac{d_j}{2}}$. Furthermore, by Equation (\ref{eq1}) and (\ref{eq8}), we obtain
$x^{(\nu+i)\frac{d_j}{2}}\beta_i(x)\in {\rm Tr}_j^{-1}\left(x^{(\nu+i)\frac{d_j}{2}}
\sum_{h=0}^{i-1}\delta^{(i)}_{(\beta_0,\beta_1,\ldots,\beta_{h})}(x)\right)$. This
is equivalent to that $\beta_i(x)\in \mathcal{W}_{(\beta_0,\beta_1,\ldots,\beta_{i-1})}^{(j,\nu; i)}$.

\par
  As stated above, according to the inductive principle we conclude that
$$\Omega_{j,\nu}=\{\sum_{i=0}^{\nu-1}\beta_i(x)f_j(x)^i\mid b_i(x)\in \mathcal{W}^{(j,\nu; i)}_{(\beta_0,\ldots,\beta_{i-2},\beta_{i-1})}, \ i=0,1,\ldots,\nu-1\}$$
in which we set $\mathcal{W}^{(j,\nu; i)}_{(\beta_0,\ldots,\beta_{i-2},\beta_{i-1})}=\mathcal{W}^{(j,\nu; 0)}$ if
$i=0$. Therefore, the number of elements in $\Omega_{j,\nu}$ is
$|\Omega_{j,\nu}|=\prod_{i=0}^{\nu-1}|\mathcal{W}^{(j,\nu; i)}_{(\beta_0,\ldots,\beta_{i-2},\beta_{i-1})}|
=(2^{m\frac{d_j}{2}})^\nu=2^{\nu m\frac{d_j}{2}}$.
\hfill $\Box$



\section{Proof of Theorem 2.6}

\noindent
In this section, we give a strict proof for Theorem 2.6 in Section 2.
To do this, we need some known results for self-dual cyclic codes  of length $2^sn$ over the ring $\mathbb{F}_{2^m}+u\mathbb{F}_{2^m}$
($u^2=0$) in \cite{s5}.

\par
  First, from \cite{s5} Theorem 5.3 and by the following substitutions:
$$p\rightarrow 2, \ \nu\rightarrow 1, \ -\nu\rightarrow 1, \
\mathcal{K}_j/\langle f_j(x)^l\rangle \rightarrow \frac{\mathbb{F}_{2^m}[x]}{\langle f_j(x)^l\rangle}
\ (1\leq l\leq 2^s-1).$$
we obtain the following conclusion for self-dual cyclic codes over $\mathbb{F}_{2^m}+u\mathbb{F}_{2^m}$.

\vskip 3mm \noindent
  {\bf Lemma 4.1}
   \textit{Using the notations above and let $x^{-1}=x^{2^sn-1}$, all distinct
self-dual cyclic codes of length $2^sn$ over $\mathbb{F}_{2^m}+u\mathbb{F}_{2^m}$ are given by
$$\mathcal{C}=\bigoplus_{j=1}^r\varepsilon_j(x)C_j \ ({\rm mod} \ x^{2^sn}-1),$$
where $C_j$ is an ideal of $\mathcal{K}_j+u\mathcal{K}_j$ given by the following two cases}:

\begin{description}
\item{($\dag$)}
   \textit{If $1\leq j\leq \rho$, $C_j$ is given by one of the following three subcases}.

\begin{description}
\item{($\dag$-1)}
   \textit{$C_j=\langle f_j(x)b(x)+u\rangle$, where
$b(x)\in f_j(x)^{2^{s-1}-1}\cdot \frac{\mathbb{F}_{2^m}[x]}{\langle f_j(x)^{2^s-1}\rangle}$ satisfying
$b(x)+\delta_jx^{-d_j}b(x^{-1})\equiv 0 \ ({\rm mod} \ f_j(x)^{2^s-1}).$}

\item{($\dag$-2)}
   $C_j=\langle f_j(x)^{2^{s-1}}\rangle$.

\item{($\dag$-3)}
    \textit{$C_j=\langle f_j(x)^{k+1}b(x)+uf_j(x)^k,f_j(x)^{k+t}\rangle$, where
$1\leq t\leq 2^s-k-1$, $1\leq k\leq 2^s-2$,
 and $b(x)\in f_j(x)^{\lceil\frac{t}{2}\rceil-1}\cdot \frac{\mathbb{F}_{2^m}[x]}{\langle f_j(x)^{t-1}\rangle}$ satisfying
$2k+t=2^s$ and}
$b(x)+\delta_jx^{-d_j}b(x^{-1})\equiv 0 \ ({\rm mod} \ f_j(x)^{t-1}).$
\end{description}

\item{($\ddag$)}
   \textit{Let $j=\rho+i$, where $1\leq i\leq \epsilon$. Then all distinct
   $N_{(2^m,d_j,2^s)}$ pairs $(C_j,C_{j+\epsilon})$ of ideals are given by the following five subcases}.

\begin{description}
\item{($\ddag$-1)}
   \textit{$C_j=\langle f_j(x)b(x)+u\rangle$ and $C_{j+\epsilon}=\langle \delta_jx^{2^sn-d_j}f_{j+\epsilon}(x)b(x^{-1})+u\rangle$
  where
$b(x)\in f_j(x)^{2^{s-1}-1}\cdot \frac{\mathbb{F}_{2^m}[x]}{\langle f_j(x)^{2^s-1}\rangle}$}.

\item{($\ddag$-2)}
   \textit{$C_j=\langle f_j(x)^{k+1}b(x)+uf_j(x)^k\rangle$ and
\begin{center}
$C_{j+\epsilon}=\langle \delta_j x^{2^sn-d_j}f_{j+\epsilon}(x)b(x^{-1})+u, f_{j+\epsilon}(x)^{2^s-k}\rangle$,
\end{center}
where $b(x)\in f_j(x)^{\lceil \frac{2^s-k}{2}\rceil-1}\cdot \frac{\mathbb{F}_{2^m}[x]}{\langle f_j(x)^{2^s-k-1}\rangle}$
 and $1\leq k\leq 2^s-1$}.

\item{($\ddag$-3)}
  \textit{$C_j=\langle f_j(x)^k\rangle$ and $C_{j+\epsilon}=\langle f_{j+\epsilon}(x)^{2^s-k}\rangle$, where $0\leq k\leq 2^s$}.

\item{($\ddag$-4)}
   \textit{$C_j=\langle f_j(x)b(x)+u,f_j(x)^{t}\rangle$ and
\begin{center}
$C_{j+\epsilon}=\langle \delta_j x^{2^sn-d_j}f_{j+\epsilon}(x)^{2^s-t+1}b(x^{-1})+uf_{j+\epsilon}(x)^{2^s-t}\rangle$,
\end{center}
where $b(x)\in f_j(x)^{\lceil\frac{t}{2}\rceil-1}\cdot \frac{\mathbb{F}_{2^m}[x]}{\langle f_j(x)^{t-1}\rangle}$ and $1\leq t\leq 2^s-1$}.

\item{($\ddag$-5)}
   \textit{$C_j=\langle f_j(x)^{k+1}b(x)+uf_j(x)^k,f_j(x)^{k+t}\rangle$ and
\begin{eqnarray*}
C_{j+\epsilon}&=&\langle \delta_j x^{2^sn-d_j}f_{j+\epsilon}(x)^{2^s-k-t+1}b(x^{-1})+uf_{j+\epsilon}(x)^{2^s-k-t},\\
&&f_{j+\epsilon}(x)^{2^s-k}\rangle,
\end{eqnarray*}
where $b(x)\in f_j(x)^{\lceil\frac{t}{2}\rceil-1}\cdot \frac{\mathbb{F}_{2^m}[x]}{\langle f_j(x)^{t-1}\rangle}$, $1\leq t\leq 2^s-k-1$ and $1\leq k\leq 2^s-2$}.
\end{description}
\end{description}

\par
  In order to prove Theorem 2.6, it is sufficiency to verify the set of codes listed by Lemma 4.1 is exactly the same as the set of codes
listed by Theorem 2.6. To do this, we need to consider the following three cases.
\begin{center}
{\bf Case 1: $j=\rho+i$ where $1\leq i\leq \epsilon$}
\end{center}

   In this case, it
is clear that the set of all pairs $(C_j,C_{j+\epsilon})$ of ideals in Case ($\ddag$) of
Lemma 4.1 is the same as the the set of all pairs $(C_j,C_{j+\epsilon})$ of ideals in Case (III)
of Theorem 2.6.
\begin{center}
{\bf Case 2: $2\leq j\leq \rho$}
\end{center}

 The ideal in Case ($\dag$-2)
of Lemma 4.1 is the same as the ideal in Case (ii-1) of
Theorem 2.6. Then we only need to consider the other two subcases:

$\diamondsuit$ Let $C_j$ be an
ideal of $\mathcal{K}_j+u\mathcal{K}_j$
in Case ($\dag$-3) of Lemma 4.1. Then
\begin{equation}
\label{eq9}
C_j=\langle f_j(x)^{k+1}b(x)+uf_j(x)^k,f_j(x)^{k+t}\rangle,
\end{equation}
where $1\leq t\leq 2^s-k-1$, $1\leq k\leq 2^s-2$,
 and $b(x)\in f_j(x)^{\lceil\frac{t}{2}\rceil-1}\cdot \frac{\mathbb{F}_{2^m}[x]}{\langle f_j(x)^{t-1}\rangle}$ satisfying
$2k+t=2^s$ and
$b(x)+\delta_jx^{-d_j}b(x^{-1})\equiv 0 \ ({\rm mod} \ f_j(x)^{t-1}).$

  As $2\leq j\leq \rho$, we have $\delta_j=1$ by Lemma 2.4(i). Let $\nu=2^{s-1}-k$. Then from $2k+t=2^s$,
$1\leq t\leq 2^s-k-1$ and $1\leq k\leq 2^s-2$, we deduce that
\begin{equation}
\label{eq10}
t=2^s-2k=2\nu, \ k=2^{s-1}-\nu \ {\rm and} \ 1\leq \nu\leq 2^{s-1}-1.
\end{equation}

\par
  Now, let $t=2\nu$ and $b(x)\in f_j(x)^{\lceil\frac{t}{2}\rceil-1}\cdot \frac{\mathbb{F}_{2^m}[x]}{\langle f_j(x)^{t-1}\rangle}
=f_j(x)^{\nu-1}\cdot \frac{\mathbb{F}_{2^m}[x]}{\langle f_j(x)^{2\nu-1}\rangle}$. As we
regard $\frac{\mathbb{F}_{2^m}[x]}{\langle f_j(x)^\nu\rangle}$ as a subset of $\frac{\mathbb{F}_{2^m}[x]}{\langle f_j(x)^{2\nu-1}\rangle}$, by the remark after Lemma 2.3 there is a unique
element $\beta(x)\in \frac{\mathbb{F}_{2^m}[x]}{\langle f_j(x)^{\nu}\rangle}$ such that
$b(x)=f_j(x)^{\nu-1}\beta(x).$
By ${\rm deg}(f_j(x))=d_j$ and Lemma 2.4(i), we have
$$f_j(x^{-1})=x^{-d_j}(x^{d_j}f_j(x^{-1}))=x^{-d_j}\widetilde{f}_j(x)
=x^{-d_j}f_j(x).$$
 This implies
$b(x^{-1})=f_j(x^{-1})^{\nu-1}\beta(x^{-1})=x^{-(\nu-1)d_j}f_j(x)^{\nu-1}\beta(x^{-1}).$
Therefore, $b(x)$ satisfies the congruence equation $b(x)+\delta_jx^{-d_j}b(x^{-1})\equiv 0$ (mod  $f_j(x)^{t-1}$),
i.e.,
$$f_j(x)^{\nu-1}\beta(x)+x^{-d_j}\cdot x^{-(\nu-1)d_j}f_j(x)^{\nu-1}\beta(x^{-1})\equiv 0 \ \ ({\rm mod} \ f_j(x)^{2\nu-1}),$$
if and only if $b(x)=f_j(x)^{\nu-1}\beta(x)$ where $\beta(x)\in \frac{\mathbb{F}_{2^m}[x]}{\langle f_j(x)^{\nu}\rangle}$
satisfies the congruence equation
$\beta(x)+x^{-\nu d_j}\beta(x^{-1})\equiv 0 \ \ ({\rm mod} \ f_j(x)^{\nu}).$

\par
  As stated above, by the definition of the set $\Omega_{j,\nu}$ and Equations (\ref{eq9}) and (\ref{eq10})
we conclude that the ideals of $\mathcal{K}_j+u\mathcal{K}_j$ in Case ($\dag$-3) of Lemma 4.1 are the
same as the ideals in Case (ii-3) of Theorem 2.6, i.e.,
$$
C_j=\langle f_j(x)^{2^{s-1}-\nu+1}b(x)+uf_j(x)^{2^{s-1}-\nu},f_j(x)^{2^{s-1}-\nu+2\nu}\rangle,
$$
where $b(x)\in f_j(x)^{\nu-1}\beta(x)$ and $\beta(x)\in \Omega_{j,\nu}$.

\par
  $\diamondsuit$ similarly, we can easily verify that the ideals
in Case ($\dag$-2) of Lemma 4.1 are the same as the ideals
in Case (ii-2) of Theorem 2.6. We omit it here.
\begin{center}
{\bf Case 3: $j=1$}
\end{center}

In this case, we have that $f_1(x)=x+1$, $\mathcal{K}_1=\frac{\mathbb{F}_{2^m}[x]}{\langle f_1(x)^{2^s}\rangle}=\frac{\mathbb{F}_{2^m}[x]}{\langle (x+1)^{2^s}\rangle}$ and
$$\frac{(\mathbb{F}_{2^m}+u\mathbb{F}_{2^m})[x]}{\langle x^{2^s}-1\rangle}
=\frac{(\mathbb{F}_{2^m}+u\mathbb{F}_{2^m})[x]}{\langle (x+1)^{2^s}\rangle}=\mathcal{K}_1+u\mathcal{K}_1.$$
Moreover,
by setting $n=1$ in Lemma 4.1, we have
  $r=\rho=1 \ {\rm and} \ \varepsilon_1(x)=1.$
From these, we deduce the following conclusion immediately.

\vskip 3mm\noindent
  {\bf Lemma 4.2} \textit{A nonempty subset $C_1$ is an ideal of the ring $\mathcal{K}_1+u\mathcal{K}_1$
listed by Cases $(\dag$-$1)$--$(\dag$-$3)$ of Lemma 4.1 if and only if $C_1$ is a self-dual cyclic code
over $\mathbb{F}_{2^m}+u\mathbb{F}_{2^m}$ of length $2^s$}.

\vskip 3mm\par
 All distinct self-dual cyclic codes
over $\mathbb{F}_{2^m}+u\mathbb{F}_{2^m}$ of length $2^s$ had been determined
by (ii)--(iv) of Theorem 2 in \cite{s8}.
From this and by Lemma 4.2, we conclude that the ideals
in Cases ($\dag$-1)--($\dag$-3) of Lemma 4.1 are the same as the ideals
in Case (I) of Theorem 2.6.

 \par
As stated above, we proved Theorem 2.6.


\section{Examples}
\noindent
 First, we consider how to calculate the number of self-dual cyclic codes over $\mathbb{F}_{2^m}+u\mathbb{F}_{2^m}$ of length
$2^sn$ from the odd positive integer $n$ directly.

\par
  Let $J_1,J_2,\ldots,J_r$ be the all distinct
$2^m$-cyclotomic cosets modulo $n$ corresponding to the factorization $x^n-1=f_1(x)f_2(x)
\ldots f_r(x)$, where $f_1(x)=x-1, f_2(x),\ldots,  f_r(x)$ are distinct monic
irreducible polynomials in $\mathbb{F}_{2^m}[x]$. Then we have $r=\rho+2\epsilon$, $J_1=\{0\}$ and

\par
 $\diamond$ $d_j=|J_j|$, $J_j=-J_j$ (mod $n$) and $2\mid d_j$, for all $j=2,\ldots,\rho$;

\par
 $\diamond$ $d_{\rho+i}=|J_{\rho+i}|=|J_{\rho+i+\epsilon}|$ and $J_{\rho+i+\epsilon}=-J_{\rho+i}$ (mod $n$), for all $1\leq j\leq \epsilon$.

\noindent
  From this and by Theorem 2.7, we can get the number of self-dual cyclic codes over $\mathbb{F}_{2^m}+u\mathbb{F}_{2^m}$ of length
$2^sn$ from the positive integer $n, s, m$ directly.

\par
  Let $m=1$. By Theorem 2.7 we list the number $N$
of self-dual cyclic codes over $\mathbb{F}_{2}+u\mathbb{F}_{2}$ of length $2^sn$,
where $s\geq 2$, $12\leq 2^sn\leq 100$ and $n\geq 3$ being odd, by the following table.
\begin{center}
{\small \begin{tabular}{l|ll|l}\hline
  $2^sn$ & $(s,n)$  &   $(r;\rho,\epsilon)$ & $N$ \\ \hline
$12$ & $(2,3)$ & $(2;2,0)$ & $49=(1+2+2^2)(1+2^{\frac{2}{2}}+2^2)$ \\
$20$ & $(2,5)$ & $(2;2,0)$ & $147=7\cdot(1+2^{\frac{4}{2}}+2^4)$ \\
$24$ & $(3,3)$ & $(2;2,0)$ & $589=(1+2+2\cdot2^2+2^3)(1+2^{\frac{2}{2}}+2^2+2^3+2^4)$ \\
$28$ & $(2,7)$ & $(3;1,1)$ & $791=7\cdot(9+5\cdot 2^3+4^3)$ \\
$36$ & $(2,9)$ & $(3;3,0)$ & $3577=7\cdot(1+2^{\frac{2}{2}}+2^2)(1+2^{\frac{6}{2}}+2^6)$ \\
$40$ & $(3,5)$ & $(2;2,0)$ & $6479=19\cdot(1+2^{\frac{4}{2}}+4^2+4^3+4^4)$ \\
$44$ & $(2,11)$ & $(2;2,0)$ & $7399=7\cdot(1+2^{\frac{10}{2}}+2^{10})$ \\
$48$ & $(4,3)$ & $(2;2,0)$ & $46501=(1+2+2\cdot2^2\cdot7+2^5)(1+\sum_{\nu=1}^8 2^\nu)$ \\
$52$ & $(2,13)$ & $(2;2,0)$ & $29127=7\cdot(1+2^{\frac{12}{2}}+2^{12})$ \\
$56$ & $(3,7)$ & $(3;1,1)$ & $139707=19\cdot(17+13\cdot 2^3+9\cdot 2^6+5\cdot 2^9+2^{12})$ \\
$60$ & $(2,15)$ & $(4;2,1)$ & $355005=49\cdot(1+2^2+2^4)\cdot(9+5\cdot 2^4+4^8)$ \\
$68$ & $(2,17)$ & $(3;3,0)$ & $521703=7\cdot(1+2^{\frac{8}{2}}+2^8)^2$ \\
$72$ & $(3,9)$ & $(3;3,0)$ & $2757109=19\cdot(2^5-1)\cdot(1+\sum_{\nu=1}^4(2^{\frac{6}{2}})^\nu)$ \\
$76$ & $(2,19)$ & $(2;2,0)$ & $1838599=7\cdot(1+2^{\frac{18}{2}}+2^{18})$ \\
$80$ & $(4,5)$ & $(2;2,0)$ & $7951671=91\cdot(1+\sum_{\nu=1}^8(2^{{\frac{4}{2}}})^\nu)$ \\
$84$ & $(2,21)$ & $(6;2,2)$ & $171508575=7\cdot7\cdot(9+5\cdot 2^3+4^3)(9+5\cdot 2^6+4^6)$ \\
$88$ & $(3,11)$ & $(2;2,0)$ & $20565619=19\cdot(1+2^{\frac{10}{2}}+2^{10}+2^{15}+2^{20})$ \\
$92$ & $(2,23)$ & $(3;1,1)$ & $29431871=7\cdot(9+5\cdot2^{11}+4^{22})$ \\
$96$ & $(5,3)$ & $(2;2,0)$ & $200669701=(1+2+8\cdot(2^7-1)+2^9)(1+\sum_{\nu=1}^{16} 2^\nu)$ \\
$100$ & $(2,25)$ & $(3;3,0)$ & $154291347=7\cdot(1+2^{\frac{4}{2}}+2^4)(1+2^{\frac{20}{2}}+2^{20})$ \\
\hline
\end{tabular} }
\end{center}

\vskip 3mm\par
 Then  we show how to list all distinct self-dual cyclic codes over $\mathbb{F}_{2^m}+u\mathbb{F}_{2^m}$ of length
$2^sn$ by use of Theorem 2.6. To save space, we only consider the case of $s=3$ and $n=3$.

In this case, we have $x^3-1=f_1(x)f_2(x)$, where
$f_1(x)=x+1 \ {\rm and} \ f_2(x)=x^2+x+1=\widetilde{f}_2(x).$
Hence $r=2$, $\rho=2$, $\epsilon=0$ and $d_2=2$.
   Using the notation in Section 2, we have
$\varepsilon_1(x)=1+x^8+x^{16} \ {\rm and} \ \varepsilon_2(x)=x^8+x^{16}.$

By use of Theorem 3.2, we calculate the sets $\Omega_{2,\nu}$, $1\leq \nu\leq 4$, and obtain:
\begin{description}
\item{}
 $\Omega_{2,1}=\{0,1+x\}$;

\item{}
 $\Omega_{2,2}=\{\beta_0(x)+\beta_1(x)f_2(x)\mid \beta_0(x)\in\{0,x\}, \ \beta_1(x)\in\{0,1\}\}$;

\item{}
 $\Omega_{2,3}=\{1+\beta_2(x)f_2(x)^2\mid \beta_2(x)\in\{1,1+x\}\}
 \cup\{1+(1+x)f_2(x)+\beta_2(x)f_2(x)^2\mid \beta_2(x)\in\{0,x\}\}
 \cup\{(1+x)f_2(x)+\beta_2(x)f_2(x)^2\mid \beta_2(x)\in\{1,1+x\}\}
 \cup \{\beta_2(x)f_2(x)^2\mid \beta_2(x)\in\{0,x\}\}$;

\item{}
$\Omega_{2,4}=\{\beta_1(x)f_2(x)+\beta_2(x)f_2(x)^2+\beta_3(x)f_2(x)^3
\mid \beta_1(x)\in \{0,x\}, \ \beta_2(x)\in \{0,1\}, \
\beta_3(x)\in \{0,1+x\}\}\\
\cup \{1+x+\beta_1(x)f_2(x)+\beta_2(x)f_2(x)^2+\beta_3(x)f_2(x)^3
 \mid \beta_1(x)\in \{1,1+x\}, \ \beta_2(x)\in \{0,1\}, \
\beta_3(x)\in \{0,1+x\}\}$.
\end{description}

\noindent
 For details of the calculation process, see Appendix A of this
paper.

\par
  By Theorem 2.6,
 all self-dual cyclic codes over $\mathbb{F}_{2}+u\mathbb{F}_{2}$ of length $24$
are given by:
$\mathcal{C}=\varepsilon_1(x)C_1\oplus \varepsilon_2(x)C_2 \ ({\rm mod} \ x^{24}-1),$
where
\begin{description}
\item{$\diamondsuit$}
  $C_1$ is one of the following $19$ ideals of the ring $\frac{\mathbb{F}_{2}[x]}{\langle(x+1)^8\rangle}+u\frac{\mathbb{F}_{2}[x]}{\langle(x+1)^8\rangle}$
  (cf. Section 5 of \cite{s8}):
 \begin{description}
\item{}
   $\langle (x+1)^4\rangle$;

\item{}
    $\langle(x+1)b(x)+u\rangle$,
    where $b(x)=b_3(x+1)^3+b_5(x+1)^5+b_6(x+1)^6$ with $b_3,b_5,b_6\in \mathbb{F}_{2}$;

\item{}
   $\langle (x+1)^4b+u(x+1)^3, (x+1)^5\rangle$ where $b\in \mathbb{F}_{2}$;

\item{}
  $\langle (x+1)^3b(x)+u(x+1)^2, (x+1)^6\rangle$, where $b(x)=b_1(x+1)+b_2(x+1)^2$ with $b_1, b_2\in \mathbb{F}_{2}$;

\item{}
  $\langle (x+1)^2b(x)+u(x+1), (x+1)^7\rangle$, where $b(x)=b_3(x+1)^3+b_4(x+1)^4$ with $b_3, b_4\in \mathbb{F}_{2}$.
\end{description}

\item{$\diamondsuit$}
 $C_2$ is one of the following $31$ ideals of the ring $\frac{\mathbb{F}_{2}[x]}{\langle f_2(x)^8\rangle}+u\frac{\mathbb{F}_{2}[x]}{\langle f_2(x)^8\rangle}$:
\begin{description}
\item{}
   $C_2=\langle f_2(x)^{4}\rangle$;

\vskip 2mm\item{}
   $C_2=\langle f_2(x)b(x)+u\rangle$, \\
   where $b(x)\in f_2(x)^{3}\cdot \Omega_{2,4}=\{f_2(x)^{3}\beta(x)\mid \beta(x)\in \Omega_{2,4}\}$;

\vskip 2mm\item{}
   $C_2=\langle f_2(x)^{5-\nu}b(x)+uf_j(x)^{4-\nu},f_j(x)^{4+\nu}\rangle$,
     where $1\leq \nu\leq 3$ and \\
     $b(x)\in f_2(x)^{\nu-1}\cdot \Omega_{2,\nu}=\{f_2(x)^{\nu-1}\beta(x)\mid \beta(x)\in \Omega_{2,\nu}\}$.
\end{description}
\end{description}

\par
  From self-dual cyclic codes $\mathcal{C}$ over $\mathbb{F}_{2}+u\mathbb{F}_{2}$ of length $24$ listed above, by the
Gray map $\phi$ from $(\mathbb{F}_{2}+u\mathbb{F}_{2})^{24}$ onto $\mathbb{F}_{2}^{48}$ defined in Section 1 one can easily obtains
$589$ self-dual and $2$-quasi-cyclic binary codes $\phi(\mathcal{C})$.

\section{Conclusions and further research}
\noindent
We have given an explicit representation for self-dual cyclic codes of length $2^sn$
over the finite chain ring $\mathbb{F}_{2^m}
+u\mathbb{F}_{2^m}$ ($u^2=0$)
and provided a clear mass formula to count the number
of all these codes. Especially, we provide an effective algorithm to
determine the key components $\Omega_{j,\nu}$ in the expression of self-dual cyclic codes of length $2^sn$
over $\mathbb{F}_{2^m}
+u\mathbb{F}_{2^m}$ by use of trace functions.

\par
   Future topics of interest include to give an explicit
representation and enumeration for repeated root self-dual cyclic codes
over $\mathbb{F}_{p^m}+u\mathbb{F}_{p^m}$ for arbitrary odd prime $p$, and to obtain some bounds for minimal distance of a self-dual and $2$-quasi-cyclic code over $\mathbb{F}_{2^m}$ derived from a self-dual cyclic code over  $\mathbb{F}_{2^m}+u\mathbb{F}_{2^m}$ by just looking at the representation of such codes.

\section*{Acknowledgments}

\noindent
Part of this work was
done when Yonglin Cao was visiting Chern Institute of Mathematics, Nankai
University, Tianjin, China. Yonglin Cao would like to thank the institution
for the kind hospitality. This research is supported in part by the National
Natural Science Foundation of China (Grant Nos. 11671235, 11801324, 61571243), the Shandong Provincial Natural Science Foundation,
China (Grant No. ZR2018BA007), the Scientific Research Fund of Hubei Provincial Key Laboratory of Applied Mathematics (Hubei University)(Grant No. AM201804) and the Scientific Research Fund of Hunan
Provincial Key Laboratory of Mathematical Modeling and Analysis in
Engineering (No. 2018MMAEZD09).

\section*{Appendix A: Calculation for the sets $\Omega_{2,\nu}$ ($1\leq \nu \leq 4$)}

\noindent
As $f_2(x)=x^2+x+1$, we have that

\begin{description}
\item{$\diamond$}
$\mathcal{F}_2=\frac{\mathbb{F}_{2}[x]}{\langle f_2(x)\rangle}=\{0,1,x,1+x\}$;
$\mathcal{H}_2=\{\xi\in \mathcal{F}_2\mid \xi^2=\xi\}=\{0,1\}=\mathbb{F}_{2}$.

\item{$\diamond$}
  Let ${\rm Tr}$ be the trace function from $\mathcal{F}_2$ onto $\mathcal{H}_2$ defined
by \\
${\rm Tr}(\xi)=\xi+\xi^2\in \mathcal{H}_2, \ \forall \xi\in \mathcal{F}_2.$

\item{$\diamond$}
  $x^{-i}\equiv x^{24-i}$ (mod $x^{24}-1$) for all integer $i$, $1\leq i\leq 23$,
and $x^3=1$ in $\mathcal{F}_2$.
\end{description}
\begin{center}
 {\bf Case $\nu=1$}
\end{center}
  By Theorem 3.2 we have
\begin{description}
\item{$\bullet$}
$\Omega_{2,1}=\mathcal{W}^{(2,1;0)}=x^{-1}\cdot {\rm Tr}^{-1}(0)=(1+x)\cdot \mathbb{F}_2=\{0,1+x\}.$
\end{description}

\begin{center}
 {\bf Case $\nu=2$}
\end{center}
    By Theorem 3.2, each element
$\beta(x)=\beta_0(x)+\beta_1(x)f_2(x)\in \Omega_{2,2}$ is determined by the following two steps:
\begin{description}
\item{\textsl{Step 2.0}}
  Let $\mathcal{W}^{(2,2;0)}=x^{-2}\cdot {\rm Tr}^{-1}(0)=\{0,x\}$.

\item{\textsl{Step 2.1}} For each $\beta_0(x)\in \mathcal{W}^{(2,2;0)}$, find an element
$\delta^{(1)}_{(\beta_0)}(x)\in \mathcal{F}_2$ satisfying
$$\beta_{0}(x)+x^{-2\cdot 2}\beta_{0}(x^{-1})=\delta^{(1)}_{(\beta_0)}(x)f_2(x)
\ ({\rm mod} \ f_2(x^2)).$$
Specifically, we have $\delta^{(1)}_{(0)}(x)=0$ and $\delta^{(1)}_{(x)}(x)=0$.
Hence
\begin{description}
\item{$\triangleright$}
$\mathcal{W}^{(2,2;1)}_{(\beta_0)}=x^{-(2+1)}\cdot {\rm Tr}^{-1}(x^{2+1}\delta^{(1)}_{(\beta_0)}(x))={\rm Tr}^{-1}(0)=\mathbb{F}_2$,
for any $\beta_0(x)\in \mathcal{W}^{(2,2;0)}$.
\end{description}

\item{}
Therefore, $\Omega_{2,2}=\{\beta_0(x)+\beta_1(x)f_2(x)\mid \beta_1(x)\in \mathcal{W}^{(2,2;1)}_{(\beta_0)},
\ \beta_0(x)\in \mathcal{W}^{(2,2;0)}\}$. Precisely, we have
\end{description}
\begin{description}
\item{$\bullet$}
$\Omega_{2,2}=\{\beta_0(x)+\beta_1f_2(x)\mid \beta_0(x)\in\{0,x\}, \ \beta_1\in\{0,1\}\}.$
\end{description}

\begin{center}
 {\bf Case $\nu=3$}
\end{center}
   By Theorem 3.2, each element
$\beta(x)=\beta_0(x)+\beta_1(x)f_2(x)+\beta_2(x)f_2(x)^2\in \Omega_{2,3}$ is determined by the following three steps:
\begin{description}
\item{\textsl{Step 3.0}}
  Let $\mathcal{W}^{(2,3;0)}=x^{-3}\cdot {\rm Tr}^{-1}(0)=\mathbb{F}_2$.

\item{\textsl{Step 3.1}} For each $\beta_0(x)\in \mathcal{W}^{(2,3;0)}$, find
elements $\delta_{(\beta_0)}^{(1)}(x),\delta_{(\beta_0)}^{(2)}(x)\in \mathcal{F}_2$ satisfying
$$\beta_{0}(x)+x^{-3\cdot 2}\beta_{0}(x^{-1})=\delta_{(\beta_0)}^{(1)}(x)f_2(x)
+\delta_{(\beta_0)}^{(2)}(x)f_2(x)^2 \ ({\rm mod} \ f_2(x)^3).$$
Specifically, we have that $\delta_{(0)}^{(1)}(x)=\delta_{(0)}^{(2)}(x)=0$;
$\delta_{(1)}^{(1)}(x)=0$ and $\delta_{(1)}^{(2)}(x)=x$.
Moreover, by $\delta_{(0)}^{(1)}(x)=\delta_{(1)}^{(1)}(x)=0$ it follows that
\begin{description}
\item{$\triangleright$}
$\mathcal{W}^{(2,3;1)}_{(\beta_0)}=x^{-(3+1)}\cdot {\rm Tr}^{-1}(x^{3+1}\delta^{(1)}_{(\beta_0)}(x))=(1+x)\cdot{\rm Tr}^{-1}(0)=\{0,1+x\}$, for any $\beta_0(x)\in \mathcal{W}^{(2,3;0)}$.
\end{description}

\item{\textsl{Step 3.2}} For each $\beta_0(x)\in \mathcal{W}^{(2,3;0)}=\{0,1\}$ and
$\beta_1(x)\in \mathcal{W}^{(2,3;1)}_{(\beta_0)}=\{0,1+x\}$, find an
element $\delta_{(\beta_0,\beta_1)}^{(2)}(x)\in \mathcal{F}_2$ satisfying
$$\beta_{1}(x)+x^{-(3+1)\cdot 2}\beta_{1}(x^{-1})+\delta^{(1)}_{(\beta_0)}(x)=\delta_{(\beta_0,\beta_1)}^{(2)}(x)f_2(x)
\ ({\rm mod} \ f_2(x)^2).$$
Specifically, we have
$\delta^{(2)}_{(0,0)}(x)=\delta^{(2)}_{(1,0)}(x)=0$ and $\delta^{(2)}_{(0,1+x)}(x)=\delta^{(2)}_{(1,1+x)}(x)$ $=x$.
Hence
\begin{description}
\item{$\triangleright$}
$\mathcal{W}^{(2,3;2)}_{(0,0)}=x^{-(3+2)}\cdot {\rm Tr}^{-1}\left(x^{3+2}\left(\delta^{(2)}_{(0)}(x)+\delta^{(2)}_{(0,0)}(x)\right)\right)
 =x\cdot{\rm Tr}^{-1}(0) \\
 =\{0,x\}$;

\item{$\triangleright$}
$\mathcal{W}^{(2,3;2)}_{(0,1+x)}=x\cdot {\rm Tr}^{-1}\left(x^{5}\left(\delta^{(2)}_{(0)}(x)+\delta^{(2)}_{(0,1+x)}(x)\right)\right)
=x{\rm Tr}^{-1}(x^5(0+x)) \\
=x\cdot{\rm Tr}^{-1}(1)=\{1+x,1\}$;

\item{$\triangleright$}
$\mathcal{W}^{(2,3;2)}_{(1,0)}=x\cdot {\rm Tr}^{-1}\left(x^{5}\left(\delta^{(2)}_{(1)}(x)+\delta^{(2)}_{(1,0)}(x)\right)\right)
 =x{\rm Tr}^{-1}(x^5(x+0)) \\
 =x\cdot{\rm Tr}^{-1}(1)=\{1+x,1\}$;

\item{$\triangleright$}
$\mathcal{W}^{(2,3;2)}_{(1,1+x)}=x\cdot {\rm Tr}^{-1}\left(x^{5}\left(\delta^{(2)}_{(1)}(x)+\delta^{(2)}_{(1,1+x)}(x)\right)\right)
=x{\rm Tr}^{-1}(x^5(x+x)) \\
=x\cdot{\rm Tr}^{-1}(0)=\{0,x\}$.
\end{description}

\item{}
Therefore, we have
\item{$\bullet$}
$\Omega_{2,3}=\left\{\beta_0(x)+\beta_1(x)f_2(x)+\beta_2(x)f_2(x)^2\mid \beta_2(x)\in \mathcal{W}^{(2,3;2)}_{(\beta_0,\beta_1)},\right. \\
\left.\beta_1(x)\in \mathcal{W}^{(2,3;1)}_{(\beta_0)},
\ \beta_0(x)\in \mathcal{W}^{(2,3;0)}\right\} \\
 =\{\beta_2(x)f_2(x)^2\mid \beta_2(x)\in\{0,x\}\} \\
 \cup \{(1+x)f_2(x)+\beta_2(x)f_2(x)^2\mid \beta_2(x)\in\{1,1+x\}\} \\
 \cup \{1+\beta_2(x)f_2(x)^2\mid \beta_2(x)\in\{1,1+x\}\} \\
 \cup\{1+(1+x)f_2(x)+\beta_2(x)f_2(x)^2\mid \beta_2(x)\in\{0,x\}\}$.
\end{description}

\begin{center}
  {\bf Case $\nu=4$}
\end{center}
  By Theorem 3.2, each element
$\beta(x)=\beta_0(x)+\beta_1(x)f_2(x)+\beta_2(x)f_2(x)^2+\beta_3(x)f_2(x)^3\in \Omega_{2,4}$ is determined by the following four steps:
\begin{description}
\item{\textsl{Step 4.0}}
  Let $\mathcal{W}^{(2,4;0)}=x^{-4}\cdot {\rm Tr}^{-1}(0)=\{0,1+x\}$.

\item{\textsl{Step 4.1}} For each $\beta_0(x)\in \mathcal{W}^{(2,4;0)}$,
find elements $\delta_{(\beta_0)}^{(1)}(x), \delta_{(\beta_0)}^{(2)}(x), \delta_{(\beta_0)}^{(3)}(x)\in \mathcal{F}_2$
satisfying
$$\beta_{0}(x)+x^{-4\cdot 2}\beta_{0}(x^{-1})=\delta_{(\beta_0)}^{(1)}(x)f_2(x)+\delta_{(\beta_0)}^{(2)}(x)f_2(x)^2
+\delta_{(\beta_0)}^{(3)}(x)f_2(x)^3$$
(mod $f_2(x)^4$).
Specifically, we have
\begin{description}
\item{$\checkmark$}
  $\delta_{(0)}^{(1)}(x)=\delta_{(0)}^{(2)}(x)=\delta_{(0)}^{(3)}(x)=0$;

\item{$\checkmark$}
  $\delta_{(1+x)}^{(1)}(x)=x$, $\delta_{(1+x)}^{(2)}(x)=1$, $\delta_{(1+x)}^{(3)}(x)=0$.
\end{description}
Hence
\begin{description}
\item{$\triangleright$}
$\mathcal{W}^{(2,4;1)}_{(0)}=x^{-(4+1)}\cdot {\rm Tr}^{-1}(x^{4+1}\delta^{(1)}_{(0)}(x))=x\cdot{\rm Tr}^{-1}(0)=\{0,x\}$;
\end{description}
\begin{description}
\item{$\triangleright$}
$\mathcal{W}^{(2,4;1)}_{(1+x)}=x^{-(4+1)}\cdot {\rm Tr}^{-1}(x^{4+1}\delta^{(1)}_{(1+x)}(x))=x\cdot{\rm Tr}^{-1}(1)=\{1,1+x\}$.
\end{description}

\item{\textsl{Step 4.2}} Let $\beta_0(x)\in \mathcal{W}^{(2,4;0)}$ and $ \beta_1(x)\in \mathcal{W}^{(2,4;1)}_{(\beta_0)}$.
Find elements $\delta^{(2)}_{(\beta_0,\beta_1)}(x)$, $\delta^{(3)}_{(\beta_0,\beta_1)}(x)\in \mathcal{F}_2$
satisfying
$$\beta_1(x)+x^{-(4+1)\cdot 2}\beta_1(x^{-1})+\delta^{(1)}_{(\beta_0)}(x)
=\delta^{(2)}_{(\beta_0,\beta_1)}(x)f_2(x)+\delta^{(3)}_{(\beta_0,\beta_1)}(x)f_2(x)^2$$
(mod $f_2(x)^3$).
Specifically, we have
\begin{description}
\item{$\checkmark$}
 $\delta^{(2)}_{(0,\beta_1)}(x)=\delta^{(3)}_{(0,\beta_1)}(x)=0$,
   $\forall \beta_1\in \mathcal{W}^{(2,4;1)}_{(0)}=\{0,x\}$;

\item{$\checkmark$}
  $\delta^{(2)}_{(1+x,\beta_1)}(x)=1$ and $\delta^{(3)}_{(1+x,\beta_1)}(x)=0$,
$\forall \beta_1\in \mathcal{W}^{(2,4;1)}_{(1+x)}=\{1,1+x\}$.
\end{description}

Hence we have the following two cases:
\begin{description}
\item{$\triangleright$}
$\mathcal{W}^{(2,4;2)}_{(0,\beta_1)}=x^{-(4+2)}\cdot {\rm Tr}^{-1}\left(x^{4+2}\left(\delta^{(2)}_{(0)}(x)+\delta^{(2)}_{(0,\beta_1)}(x)\right)\right)
={\rm Tr}^{-1}(0)=\{0,1\}$, for any $\beta_1(x)\in \mathcal{W}^{(2,4;1)}_{(0)}$.

\item{$\triangleright$}
$\mathcal{W}^{(2,4;2)}_{(1+x,\beta_1)}=x^{-(4+2)}\cdot {\rm Tr}^{-1}\left(x^{4+2}\left(\delta^{(2)}_{(1+x)}(x)+\delta^{(2)}_{(1+x,\beta_1)}(x)\right)\right)
={\rm Tr}^{-1}(0)$ $=\{0,1\}=\mathcal{W}^{(2,4;2)}_{(0,\beta_1)}$, for any $\beta_1(x)\in \mathcal{W}^{(2,4;1)}_{(1+x)}$.
\end{description}

\item{\textsl{Step 4.3}} Let $\beta_0(x)\in \mathcal{W}^{(2,4;0)}$, $\beta_1(x)\in \mathcal{W}^{(2,4;1)}_{(\beta_0)}$ and $\beta_2(x)\in \mathcal{W}^{(2,4;2)}_{(\beta_0,\beta_1)}$.
Find an element $\delta_{(\beta_0,\beta_1,\beta_2)}^{(3)}(x)\in\mathcal{F}_2$ satisfying
$$\beta_2(x)+x^{-(4+2)\cdot 2}b_2(x^{-1})+\delta_{(\beta_0)}^{(2)}(x)+\delta_{(\beta_0,\beta_1)}^{(2)}(x)
=\delta_{(\beta_0,\beta_1,\beta_2)}^{(3)}(x)f_2(x)$$
(mod $f_2(x)^2$).
Specifically, for any $\beta_0(x)\in \mathcal{W}^{(2,4;0)}$ and $\beta_1(x)\in \mathcal{W}^{(2,4;1)}_{(\beta_0)}$ we have
that $\delta_{(\beta_0,\beta_1,0)}^{(3)}(x)=\delta_{(\beta_0,\beta_1,1)}^{(3)}(x)=0$. Hence
\begin{description}
\item{$\triangleright$}
$\mathcal{W}^{(2,4;3)}_{(\beta_0,\beta_1,\beta_2)}
=x^{-(4+3)}\cdot {\rm Tr}^{-1}(x^{4+3}(\delta^{(3)}_{(\beta_0)}(x)+\delta^{(3)}_{(\beta_0,\beta_1)}(x)
+\delta^{(3)}_{(\beta_0,\beta_1,\beta_2)}(x)))
$ $=(1+x)\cdot{\rm Tr}^{-1}(0)=\{0,1+x\}$, \\
for any $\beta_0(x)\in \mathcal{W}^{(2,4;0)}$, $\beta_1(x)\in \mathcal{W}^{(2,4;1)}_{(\beta_0)}$
and $\beta_2(x)\in \mathcal{W}^{(2,4;2)}_{(\beta_0,\beta_1)}$.
\end{description}

\item{}
As stated above, we have
\begin{eqnarray*}
\Omega_{2,4}&=&\{\beta_0(x)+\beta_1(x)f_2(x)+\beta_2(x)f_2(x)^2+\beta_3(x)f_2(x)^3 \\
 && \ \  \mid
\beta_3(x)\in \mathcal{W}^{(2,4;3)}_{(\beta_0,\beta_1,\beta_2)}, \ \beta_2(x)\in \mathcal{W}^{(2,4;2)}_{(\beta_0,\beta_1)}, \ \beta_1(x)\in \mathcal{W}^{(2,4;1)}_{(\beta_0)}, \\
&& \ \ \ \beta_0(x)\in \mathcal{W}^{(2,4;0)}\} \\
&=&\{\beta_1(x)f_2(x)+\beta_2(x)f_2(x)^2+\beta_3(x)f_2(x)^3 \\
 && \ \  \mid
\beta_3(x)\in \mathcal{W}^{(2,4;3)}_{(0,\beta_1,\beta_2)}, \ \beta_2(x)\in \mathcal{W}^{(2,4;2)}_{(0,\beta_1)}, \ \beta_1(x)\in \mathcal{W}^{(2,4;1)}_{(0)}\}\\
&&\cup\{1+x+\beta_1(x)f_2(x)+\beta_2(x)f_2(x)^2+\beta_3(x)f_2(x)^3 \\
 && \ \  \mid
\beta_3(x)\in \mathcal{W}^{(2,4;3)}_{(1+x,\beta_1,\beta_2)}, \ \beta_2(x)\in \mathcal{W}^{(2,4;2)}_{(1+x,\beta_1)}, \ \beta_1(x)\in \mathcal{W}^{(2,4;1)}_{(1+x)}\}.
\end{eqnarray*}

Precisely, we have
\end{description}
\begin{description}
\item{$\bullet$}
$\Omega_{2,4}=\{\beta_1(x)f_2(x)+\beta_2(x)f_2(x)^2+\beta_3(x)f_2(x)^3
\mid \beta_1(x)\in \{0,x\}, \ \beta_2(x)\in \{0,1\}, \
\beta_3(x)\in \{0,1+x\}\}\\
\cup \{1+x+\beta_1(x)f_2(x)+\beta_2(x)f_2(x)^2+\beta_3(x)f_2(x)^3
 \mid \beta_1(x)\in \{1,1+x\}, \ \beta_2(x)\in \{0,1\}, \
\beta_3(x)\in \{0,1+x\}\}$.
\end{description}


\end{document}